%% file: main.tex
\documentclass[12pt]{article}
\usepackage[a4paper,top=2cm,bottom=2cm,left=1.5cm,right=1.5cm,marginparwidth=1.75cm]{geometry}
\usepackage{amsmath,amsfonts,amssymb,amsthm,graphicx,booktabs,url,authblk,bm}
\usepackage[authoryear,longnamesfirst,sort]{natbib}
\usepackage{setspace,longtable,array,float,caption}
\usepackage[hidelinks]{hyperref}

\input{def}
\newtheorem{theorem}{Theorem}
\title{Functional meta-analysis of survival curves across reporting formats and time points}
\author[1]{Keisuke Hanada\thanks{Department of Biostatistics, Wakayama Medical University, Kimiidera, Wakayama, 641-8509, Japan \quad E-mail: \href{mailto:khanada@wakayama-med.ac.jp}{khanada@wakayama-med.ac.jp}}}
\affil[1]{Department of Biostatistics, Faculty of Medicine, Wakayama Medical University}
\date{}
\hypersetup{pdftitle={Functional meta-analysis of survival curves across reporting formats and time points},pdfauthor={Keisuke Hanada}}
\begin{document}
\maketitle
\begin{abstract}
Meta-analyses of survival outcomes commonly pool log hazard ratios or survival probabilities at selected time points. When reporting formats and observation times differ across studies, methods are needed to use available survival information for estimation at a common target time. We propose a functional random-effects model on the complementary log--log scale that synthesizes Kaplan--Meier curves and survival probabilities at discrete times as functional data. Studies that do not report survival at the target time contribute to its estimation through the use of their observed curves to estimate group-specific mean functions. The estimated between-group contrast is interpreted as a log hazard ratio under proportional hazards and as a time-specific log cumulative hazard ratio under non-proportional hazards. In simulations with delayed effects and crossing survival curves, the root mean squared error for the log cumulative hazard ratio was lower than that of the two-stage Cox method, and interval coverage was close to the nominal level. In an application to disease-free survival in resectable non-small cell lung cancer, the two-stage Cox method estimated an average hazard ratio, whereas the proposed method captured nonlinear changes in the cumulative hazard ratio over time.
\end{abstract}
\noindent\textit{Key words and phrases:} Cumulative hazard ratio; Functional data analysis; Kaplan--Meier curve; Meta-analysis; Non-proportional hazards.
\setstretch{1.3}

\section{Introduction}
\label{sec:intro}

Meta-analyses of survival outcomes pool study-specific log hazard ratios or survival probabilities at selected times. Log hazard ratios are obtained mainly from Cox regression in each study. When the log hazards ratios are unreported, their estimates and variances can be derived from published summary information \citep{tierney2025updated}. Synthesizing survival probabilities, such as 3-year survival, summarizes outcomes at clinically relevant times. When proportional hazards (PH) does not hold, summaries capturing time-varying between-group differences are useful. Under non-proportional hazards (NPH), restricted mean survival time or the average hazard over a specified evaluation period can be used \citep{royston2013restricted,wei2015meta,uno2023ratio}. Generalized linear mixed model (GLMM) meta-analyses that directly specify the outcome distribution have also been proposed \citep{hanada2026glmm} and can be applied to survival outcomes by specifying a parametric distribution. In each case, synthesis depends on study-level information suited to the estimand and model.

Even when a required summary measure is unreported, a Kaplan--Meier (KM) curve or survival probabilities at discrete times may be available. KM curves may include numbers at risk at only a few times, and survival probabilities and confidence intervals may be reported at times other than the meta-analysis target. For example, in a meta-analysis of 3- or 5-year survival in non-small cell lung cancer, studies reporting KM curves were excluded because numbers censored and loss to follow-up could not be clearly determined \citep{lee2018location}. In a meta-analysis of metastatic colorectal cancer, hazard ratios for primary-tumor laterality were unreported in 33 eligible studies, and KM curves yielded additional estimates for 8 of them \citep{vanderkruijssen2023time}. A methodological study of oncology papers reporting long-term outcomes also found that using event counts, KM curves, and fixed-time survival probabilities increased available hazard ratio estimates from 108 directly reported estimates to 240 \citep{hebert2022methodology}.

Methods have been developed to reconstruct individual participant data from KM curve images \citep{guyot2012enhanced,liu2021ipdfromkm}, but reconstruction requires assumptions about censoring. Moreover, reducing curves to survival probabilities at a few times prevents direct use of the curve shape between these time points. A distribution-free method has been proposed to pool survival curves from single-arm studies \citep{combescure2014meta}, as have methods that pool survival probabilities at multiple times while accounting for within-study correlations \citep{dear1995,arends2008}. Using survival estimates at many closely spaced time points, however, requires handling large covariance matrices with strong within-study correlations, which can make computation demanding and numerically unstable. Methods treating fixed-time survival probabilities as multivariate outcomes also require more mean and between-study covariance parameters as the number of time points increases \citep{jackson2014multivariate}.

We propose a functional random-effects model on the complementary log--log (cloglog) scale to synthesize different forms of time information. The model treats curves and survival information at discrete times as common functional inputs, allowing studies that do not report the time of interest to contribute to inference without reconstructing individual participant data. We estimate group-specific mean functions and between-study variation by drawing on the treatment of means and random effects as functions \citep{guo2002functional}, functional mixed models allowing irregular and sparse observation grids \citep{cederbaum2016functional}, and estimation with finite bases and mixed models \citep{scheipl2015functional}. For example, when estimating 36-month survival, a curve observed only through 24 months contributes survival probabilities and the shape of the between-group difference over its observed interval to mean-function estimation; evaluating that function at 36 months then allows the shorter curve to contribute to estimation at the target time.

This study makes three main contributions: (i) synthesizing survival information from studies with different reporting formats and observation times within a common framework to estimate survival curves and time-varying between-group differences; (ii) allowing studies that do not directly report the target time to contribute through their curves, increasing the number of studies available for meta-analysis; and (iii) using survival functions as observations to extend synthesis of log hazard ratios under PH to inference on time-specific log cumulative hazard ratios under NPH. These contributions broaden the utility of meta-analysis for survival outcomes.

Section~\ref{sec:model} presents the motivating example, estimands, and existing methods. Section~\ref{sec:method} describes the construction of observed information, estimation procedure, and statistical properties of the proposed method. Section~\ref{sec:simulation} evaluates finite-sample performance through simulations, and Section~\ref{sec:rda} presents a real-data application. Section~\ref{sec:discussion} discusses the implications of this study and future work.

\section{Problem formulation}
\label{sec:model}

Let $T$ denote event time, $C$ censoring time, $Z$ study membership, and $G$ treatment group or a prespecified subgroup. Define the survival function for study $k=1,\ldots,K$ and group $g=1,\ldots,G$ as $S_{kg}(t)=\Pr(T>t\mid Z=k,G=g)$. We assume independent observations across individuals and censoring independent conditional on $(Z,G)$.

\subsection{Motivating example and available information}
\label{sec:motivating-example}

As a motivating example, we consider a meta-analysis of disease-free survival (DFS) in resectable non-small cell lung cancer \citep{chen2023prognostic}. Groups are defined by positive or negative baseline circulating tumor DNA (ctDNA). Using publicly available data, we include 12 studies with observed baseline ctDNA status for which survival information can be constructed for both groups. The primary estimand is the 36-month cumulative hazard ratio for the positive versus negative group. Follow-up duration and event-time distributions differ across studies. Patient-level data are available for 10 studies; the other 2 contribute only KM curves from the original articles and limited risk-set information.

\begin{figure}[htbp]

    \centering
    \includegraphics[width=0.98\linewidth]{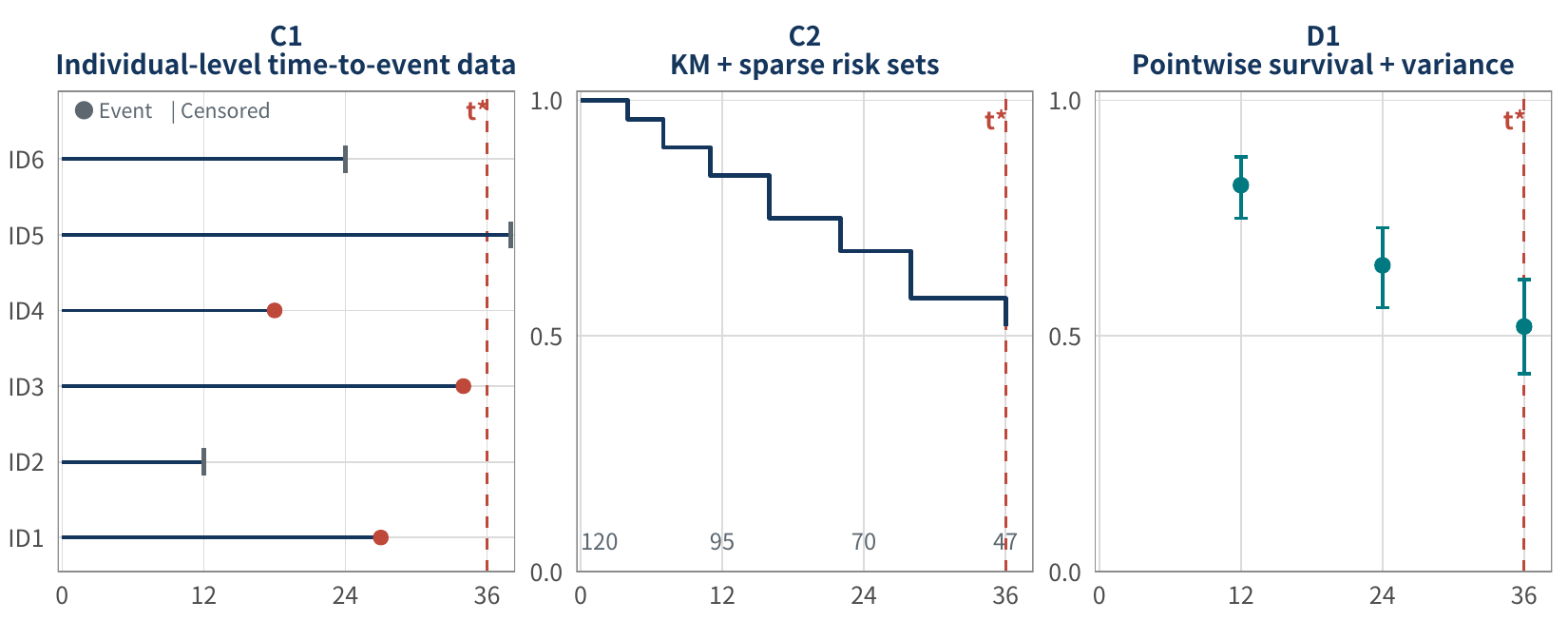}
    \caption{Schematic illustration of survival information in different reporting formats. The dashed line indicates the prespecified target time $t^*$. Values in the figure are for illustration.}
    \label{fig:information-flow}

\end{figure}

As in this example, available survival information may differ across studies. Information formats can be classified as (C1) complete curve information, (C2) curve information with sparse risk sets, or (D1) discrete-time information with marginal variance (Figure~\ref{fig:information-flow}). The motivating example includes 10 C1 studies and 2 C2 studies, with no D1 studies. However, studies with information such as D1 may be excluded from a meta-analysis because, for example, hazard ratios are not reported. For evaluation times $t_{kg1} < \dots < t_{kgn_{kg}}$ in study $k$, group $g$, denote the estimated survival function and its marginal variance by
\begin{align*}
    \widehat{S}_{kgj} &= \widehat{S}_{kg}(t_{kgj}), \quad
    \widehat\sigma^2_{kgj} = \mathrm{Var}(\widehat{S}_{kgj}).
\end{align*}
For C1, KM curves and Greenwood variances can be constructed from individual participant data or similar sources; for C2, variances can be approximated from reconstructed curves and reported numbers at risk. D1 provides survival probabilities and variance information at reported times. We convert observations from each format to the common input $\{t_{kgj},\widehat S_{kgj},\widehat\sigma^2_{kgj}\}_{j=1}^{n_{kg}}$ (Table~\ref{tab:available-summary-information}).

\begin{table}[tp]
\centering
\caption{Comparison of survival information available in C1, C2, and D1}
\label{tab:available-summary-information}
\small
\setlength{\tabcolsep}{4pt}
\begin{tabular}{p{0.14\textwidth}p{0.26\textwidth}p{0.27\textwidth}p{0.25\textwidth}}
\toprule
Common input & C1 & C2 & D1 \\
\midrule
$t_{kgj}$ & Evaluation times of KM curve & Evaluation times of image-reconstructed curve & Reported times \\
$\widehat S_{kgj}$ & KM estimate & Survival probability reconstructed from image & Reported survival probability \\
$\widehat\sigma^2_{kgj}$ & Greenwood variance & Reference variance approximated from sparse risk sets & Reported SE, confidence interval, or working approximation \\
\bottomrule
\end{tabular}

\end{table}

\subsection{Model and estimands}

The cloglog transformation of the survival function equals the log cumulative hazard, so between-group differences on this scale can be interpreted as log cumulative hazard ratios. We decompose the transformed survival function into a group-specific mean time trend and a study-specific deviation, and assume the functional random-effects model
\begin{align}
\eta_{kg}(t)
&=
\log\{-\log S_{kg}(t)\}
=
o_\rho(t)+\mu_g(t)+b_{kg}(t).
\label{eq:eta-model}
\end{align}
Here, $t_0>0$ is a reference time, $\rho>0$ is a constant, and $o_\rho(t)=\rho\log(t/t_0)$ is a known offset. The group-specific mean function excluding the offset is $\mu_g(t)$, and $b_{kg}(t)$ is a between-study random effect represented by a Gaussian process with mean $0$ and covariance function $\Gamma_{kg}(t,t')$. The offset allows the meta-analytic survival function below to extend continuously to the origin and satisfy $S_g^{\mathrm{meta}}(0)=1$.

The estimand is the group-specific mean function $\mu_g$, from which group-specific survival functions, cumulative hazard ratios, and survival probability differences follow. The common offset cancels in a between-group contrast, so the log cumulative hazard ratio for groups $g_1$ and $g_2$ is
\begin{align*}
\Delta_\mu(t; g_1, g_2)
&=
\mu_{g_1}(t)-\mu_{g_2}(t)
=\log\left\{\frac{H_{g_1}^{\mathrm{meta}}(t)}{H_{g_2}^{\mathrm{meta}}(t)}\right\}.
\end{align*}
Here, $H_g^{\mathrm{meta}}(t)=\exp\{o_\rho(t)+\mu_g(t)\}$. The group-$g$ meta-analytic survival function is defined as
\begin{align*}
S_g^{\mathrm{meta}}(t)
&=\exp\{-H_g^{\mathrm{meta}}(t)\}.
\end{align*}
The contrast $\Delta_\mu(t; g_1, g_2)$ is a time-constant log hazard ratio under PH and a log cumulative hazard ratio up to time $t$ under NPH. Applying the inverse cloglog transformation to $o_\rho(t)+\mu_g(t)$ for each group gives survival probabilities and their difference $\Delta_S(t;g_1,g_2)=S_{g_1}^{\mathrm{meta}}(t)-S_{g_2}^{\mathrm{meta}}(t)$. Its integral over a prespecified period $[0,\tau]$, $\int_0^\tau\Delta_S(t;g_1,g_2)\,dt$, is the between-group difference in restricted mean survival time. The quantity $S_g^{\mathrm{meta}}(t)$ is the survival function for a hypothetical study whose random effect satisfies $b_{kg}(t)=0$. We refer to such a hypothetical study as a \emph{typical study}. We next describe existing methods in relation to these estimands.

\subsection{Existing synthesis methods}
\label{sec:existing-methods}

\paragraph{Two-stage Cox method}
Under PH, $\Delta_\mu(t;g_1,g_2)$ is a time-constant mean log hazard ratio, abbreviated below as $\Delta_\mu$. The two-stage Cox method fits a Cox regression \citep{cox1972regression} with group as a covariate in each study to obtain the study-specific estimated log hazard ratio $\widehat\Delta_{\mu,k}$ and estimated sampling variance $s_k^2$. It approximates each estimate's sampling distribution by a normal distribution and assumes the study-specific log hazard ratios follow a normal distribution with mean $\Delta_\mu$ \citep{higgins2009re}. If $\widehat\tau_{\mathrm{HR}}^2$ denotes the estimated between-study variance, the pooled estimator is
\begin{align*}
\widehat\Delta_\mu
&=\frac{\sum_{k=1}^{K}
\widehat\Delta_{\mu,k}/(s_k^2+\widehat\tau_{\mathrm{HR}}^2)}
{\sum_{k=1}^{K}1/(s_k^2+\widehat\tau_{\mathrm{HR}}^2)}.
\end{align*}
This method summarizes a time-constant between-group effect while accounting for within-study variance and between-study heterogeneity. Under NPH, a study-specific Cox coefficient reduces a time-varying difference to one value; its pooled estimate therefore has a different estimand from the time-specific log cumulative hazard ratio $\Delta_\mu(t;g_1,g_2)$ in Section 2.2.

\paragraph{Arends method}
A multivariate random-effects model using survival probabilities and standard errors at a finite set of analysis times has also been proposed \citep{arends2008}. Let $\widehat{\boldsymbol S}_k$ be the vector of survival probabilities across groups and times in study $k$. The method fits a linear mixed model after applying the cloglog transformation componentwise:

\begin{align*}
\log\{-\log\widehat{\boldsymbol S}_k\}
&=\X_k\boldsymbol\alpha+\Z_k\u_k+\boldsymbol\varepsilon_k,
\quad \u_k\sim N(\boldsymbol 0,\D),\quad
\boldsymbol\varepsilon_k\sim N(\boldsymbol 0,\Sigma_k)
\end{align*}

Here, $\X_k$ is the design matrix for time or log time, group, and their interactions; $\boldsymbol\alpha$ is the fixed-effect coefficient vector; and $\Z_k$ is the random-effects design matrix. The independent terms $\u_k$ and $\boldsymbol\varepsilon_k$ have between-study covariance $\D$ and within-study covariance $\Sigma_k$ at the selected times, respectively. They construct $\Sigma_k$ from transformed marginal standard errors and correlations across time based on survival probabilities, then iterate between fitting the mixed model and updating correlations using predicted survival probabilities. Treating time as continuous allows us to combine observations reported at different times across studies and estimate curves between the observed time points. If the mean curve follows the specified regression model, the group-specific curves and between-group cloglog differences from the fixed effects correspond to the estimands in Section 2.2.

Applying the Arends method to dense KM curves requires within-study covariance matrices for many evaluation times and estimation with correlation updates. Restricting the selected times prevents use of survival probabilities at unselected times. Mixed curve and discrete-time reports also differ in evaluation-point locations and density. A method is therefore needed that reflects the curves' observed intervals in the estimation criterion while limiting reliance on constructing and updating sampling correlation matrices.

\section{Proposed method}
\label{sec:method}

This section presents a method that combines dense curves such as C1/C2 and sparse time-point information such as D1 through a common mean function, allowing studies that do not directly report the target time to contribute to estimation.

\subsection{Working model based on basis expansions}
\label{sec:working-model}

We represent the functions in model~\eqref{eq:eta-model} using basis expansions so that studies with different observation times can contribute to common coefficients. For $\mu_g(t)$ and $b_{kg}(t)$, respectively, we use fixed basis vectors $\x_g(t)$ and $\z_g(t)$ of dimensions $p_1$ and $p_2$, and approximate
\begin{align}
\label{eq:eta-basis-expansion}
\widetilde{\eta}_{kg}(t) &= o_\rho(t)+\widetilde{\mu}_g(t)+\widetilde{b}_{kg}(t).
\end{align}
Here, $\widetilde{\mu}_g(t)=\x_g(t)^\top\bbe$ and $\widetilde{b}_{kg}(t)=\z_g(t)^\top\u_k$, where $\u_k$ follows a multivariate normal distribution with mean $\boldsymbol{0}$ and covariance matrix $\D$. We use cubic B-spline bases for the group-specific mean functions \citep{scheipl2015functional}.

Basis approximation error arises in both the mean function and between-study variation, so inference for the true function requires its evaluation together with stochastic coefficient-estimation error. Sieve $M$-estimation provides asymptotic properties accounting for function approximation error by increasing the basis dimension with the number of studies \citep{chen2007large}. Here, we fix the basis dimension and estimate the working-model coefficient $\bbe$ and between-study covariance $\D$. Inference for the pseudo-true value of this working model is given in Section~\ref{sec:asymptotic-inference}.

\subsection{Construction of observed information}
\label{sec:information-specific-estimates}

From each information format in Table~\ref{tab:available-summary-information}, we construct the common observed information $\{t_{kgj},\widehat S_{kg}(t_{kgj}),\widehat\sigma^2_{kg}(t_{kgj})\}_{j=1}^{n_{kg}}$.

For C1, we calculate the KM estimate \citep{kaplan1958nonparametric} and Greenwood marginal variance from individual-level data. Let $Y_{kg}(u)$ be the number at risk immediately before event time $u$ and $d_{kg}(u)$ the number of events at that time in group $g$ of study $k$. Then
\begin{align*}
\widehat S_{kg}(t)
&=\prod_{u\leq t}\left\{1-\frac{d_{kg}(u)}{Y_{kg}(u)}\right\},\\
\widehat\sigma^2_{kg}(t)
&=\widehat S_{kg}(t)^2\sum_{u\leq t}
\frac{d_{kg}(u)}{Y_{kg}(u)\{Y_{kg}(u)-d_{kg}(u)\}}.
\end{align*}

For C2, only an image of the KM curve and risk sets at a few times are available. We denote the KM curve reconstructed from the image by $\widehat S_{kg}(t)$ and define its decrement rate at a time $u$ of decrease as $\widehat q_{kg}(u)=1-\widehat S_{kg}(u)/\widehat S_{kg}(u^-)$, where $u^-$ is immediately before $u$. Using the KM identity $\widehat q_{kg}(u)=d_{kg}(u)/Y_{kg}(u)$, each summand in the Greenwood variance can be written as $\widehat q_{kg}(u)/[Y_{kg}(u)\{1-\widehat q_{kg}(u)\}]$. At each event time, we approximate the number at risk by carrying forward the most recently reported value, $\widetilde Y_{kg}^{\mathrm{ref}}(u)$. The reference Greenwood marginal variance is
\begin{align*}
\widehat\sigma_{kg}^2(t)
&=\widehat S_{kg}(t)^2\sum_{u\leq t}
\frac{\widehat q_{kg}(u)}
{\widetilde Y_{kg}^{\mathrm{ref}}(u)\{1-\widehat q_{kg}(u)\}}.
\end{align*}
We use the resulting variance estimate as the working variance for pooled estimation. If the number at risk decreases between reported times, carrying values forward may underestimate the marginal variance. The variance of the coefficient estimator obtained using this working variance is assessed with the study-level sandwich variance estimator.

For D1, we use the survival probability $\widehat S_{kg}(t_{kgj})$ at reported time $t_{kgj}$ and its marginal variance $\widehat\sigma^2_{kg}(t_{kgj})$. We obtain the variance by squaring a reported standard error or by using the scale and construction of a reported confidence interval. For example, a symmetric $(1-\alpha)$ normal-approximation confidence interval $[\widehat S_{kg}^{\mathrm L}(t_{kgj}),\widehat S_{kg}^{\mathrm U}(t_{kgj})]$ on the survival-probability scale gives
\begin{align*}
\widehat\sigma^2_{kg}(t_{kgj})
&\approx\left\{\frac{\widehat S_{kg}^{\mathrm U}(t_{kgj})-\widehat S_{kg}^{\mathrm L}(t_{kgj})}{2z_{1-\alpha/2}}\right\}^2,
\end{align*}
where $z_{1-\alpha/2}$ is the $1-\alpha/2$ standard normal quantile.

\subsection{Estimation}
\label{sec:estimation}

We apply the cloglog transformation to the survival estimates and subtract the known offset to obtain the working observation model corresponding to the basis expansion in~\eqref{eq:eta-basis-expansion}:

\begin{align*}
    y_{kg}^{(\rho)}(t) &= \log\{-\log \widehat S_{kg}(t)\} - o_\rho(t) = \x_g(t)^\top \bbe + \z_g(t)^\top \u_k + \varepsilon_{kg}(t).
\end{align*}

Assuming that $\varepsilon_{kg}(t)$ has mean zero, we approximate its variance by the delta method:
\begin{align*}
v_{kg}^2(t) &=
\mathrm{Var}(\varepsilon_{kg}(t))
\approx
\frac{\hat\sigma^2_{kg}(t)}
{\hat S_{kg}(t)^2\{\log\hat S_{kg}(t)\}^2}.
\end{align*}
To prevent a study's contribution to the residual sum of squares from increasing merely because the same curve is evaluated at more time points, we supplement inverse-variance weighting with time-interval weights. Let the evaluation period be $\mathcal T=[0,\tau]$, and assign each observation time an interval $I_{kgj}=[a_{kgj},b_{kgj})$; the final interval includes its right endpoint. For C1 and C2, we valuate the curve at the midpoint of each interval over which the KM curve is constant.
For D1, we evaluate at reported times and use the midpoints of adjacent reported times as boundaries. The right endpoint of the final D1 interval is the final reported time, so the weights cover only the range through that time. We define $w_{kgj}=(b_{kgj}-a_{kgj})/\tau$ and weight each squared residual by the product of this time-interval weight and the inverse variance.

To limit excessive curvature, we penalize the second derivative of the mean function. Define the penalty matrix as
\begin{align*}
(\P)_{ij}
&=
\sum_{g=1}^{G}\int_0^\tau
\bigl[\x_g''(t)\bigr]_i
\bigl[\x_g''(t)\bigr]_j\,dt,
\quad i,j=1,\ldots,\dim(\bbe),
\end{align*}
and let the smoothing parameter be $\lambda>0$. Adding the quadratic penalty corresponding to the normal random-effects distribution and eliminating $\u_k$ gives the penalized quadratic criterion

\begin{align*}
\ell^w(\bbe, \D, \lambda) &= \sum_{k=1}^K (\y_{k}^{(\rho)} - \X_k \bbe)^\top \V_k^{-1} (\y_{k}^{(\rho)} - \X_k \bbe) + \lambda\bbe^\top\P\bbe,
\end{align*}

where $\V_k=\Z_k\D\Z_k^\top+\W_k^{-1/2}\Omega_k\W_k^{-1/2}$. The first term in the criterion measures fit to observations across studies, and the second penalizes roughness of the mean function. The vector $\y_k^{(\rho)}$ stacks $y_{kg}^{(\rho)}(t_{kgj})$ by group $g=1,\ldots,G$ and then by time $j=1,\ldots,n_{kg}$ within group; $\X_k$ and $\Z_k$ have rows $\x_g(t_{kgj})^\top$ and $\z_g(t_{kgj})^\top$ in that order. In the same order, $\W_k$ and $\Omega_k$ place the time-interval weights $w_{kgj}$ and delta-method variances $v_{kg}^2(t_{kgj})$ on their diagonals. We assume working independence across times in $\Omega_k$. The two terms in $\V_k$ represent between-study variation and within-study working variance adjusted by time-interval weights, respectively. Coefficients are estimated using this working covariance, while their uncertainty is assessed using the study-level variance estimator in the next section.

For fixed $\D$ and $\lambda$, the unconstrained coefficient estimator minimizing $\ell^w$ is

\begin{align*}
\widetilde\bbe(\D,\lambda)
&=\left(\sum_{k=1}^K\X_k^\top\V_k^{-1}\X_k+\lambda\P\right)^{-1}
\sum_{k=1}^K\X_k^\top\V_k^{-1}\y_k^{(\rho)}.
\end{align*}

We substitute this coefficient solution and select the variance component $\D$ and smoothing parameter $\lambda$ by minimizing the following restricted maximum likelihood (REML)-type pseudo-likelihood criterion:

\begin{align*}
\ell_{\mathrm{REML}}(\D,\lambda)
={}&\ell^w\{\widetilde\bbe(\D,\lambda),\D,\lambda\}
+\sum_{k=1}^K\log|\V_k|\\
&+\log\left|\sum_{k=1}^K\X_k^\top\V_k^{-1}\X_k+\lambda\P\right|
-r_{\P}\log\lambda,
\end{align*}

where $r_{\P}=\operatorname{rank}(\P)$. Finally, we impose the condition for nondecrease of the cumulative hazard $H_g^{\mathrm{meta}}(t)$, $\rho+t\x_g'(t)^\top\bbe\geq0$, on a prespecified grid $\mathcal G$ with $(\widehat\D,\widehat\lambda)$ fixed, and estimate $\bbe$ by

\begin{align*}
    \widehat \bbe &= \arg\min_{\bbe} \, \ell^w(\bbe, \widehat \D,\widehat \lambda) \quad \text{subject to} \quad \rho + t \x_g'(t)^\top \bbe \geq 0, \; \forall g,\ t\in\mathcal G.
\end{align*}

The smoothing parameter can also be selected by study-level cross-validation. Setting $\lambda=0$ allows estimation without a smoothing penalty. In this case, we remove the roughness penalty and the term $-r_{\P}\log\lambda$ from the REML-type criterion when selecting $\D$. Removing the penalty makes changes in curve shape easier to represent but increases estimation variance and the potential for overfitting.

\subsection{Asymptotic normality and interval estimation}
\label{sec:asymptotic-inference}

Fix basis dimension $p_1$ and number of groups $G$, and let the number of independent studies satisfy $K\to\infty$. If $\V_k^{\star}$ is the limit of the working covariance matrix $\V_k$, the pseudo-true coefficient for the fixed basis is defined as
\begin{align}
\bbe^\star
&=\arg\min_{\bbe}
E\!\left\{(\y_k^{(\rho)} - \X_k\bbe)^\top
(\V_k^{\star})^{-1}(\y_k^{(\rho)} - \X_k\bbe)\right\},
\label{eq:pseudo-true-target}
\end{align}
and $\mu_g^\star(t)=\x_g(t)^\top\bbe^\star$ is the best approximation to the group-specific mean function under this population criterion. Define the study-specific score by

\begin{align*}
\boldsymbol\psi_k^\star &= \X_k^\top (\V_k^\star)^{-1} (\y_k^{(\rho)} - \X_k\bbe^\star), \quad
\H = E[\X_k^\top (\V_k^\star)^{-1} \X_k], \quad
\J = E[\boldsymbol\psi_k^\star\boldsymbol\psi_k^{\star\top}].
\end{align*}

Survival probabilities at multiple times within the same study are correlated. We therefore aggregate all observations within each study into a study-specific score and estimate the sandwich variance with studies as independent units, reflecting within-study correlations in the variance of the coefficient estimator \citep{liang1986longitudinal}. Let $\widehat{\boldsymbol\psi}_k=\X_k^\top \V_k^{-1} (\y_k^{(\rho)} - \X_k\widehat\bbe)$. The study-level sandwich variance estimator is

\begin{align}
\widehat\V_{\mathrm{sand}}
&=\widehat\A^{-1}
\left(\sum_{k=1}^K
\widehat{\boldsymbol\psi}_k\widehat{\boldsymbol\psi}_k^\top\right)
\widehat\A^{-1},
\quad
\widehat\A = \sum_{k=1}^K\X_k^\top \V_k^{-1} \X_k + \widehat\lambda\P.
\label{eq:cluster-sandwich}
\end{align}

\begin{theorem}[Asymptotic properties]
\label{thm:asymptotic-normality}
Suppose the regularity conditions in Appendix B.1 hold. Then $\widehat \bbe$ converges in probability to $\bbe^\star$, and
\begin{align*}
\sqrt K(\widehat\bbe-\bbe^\star)
&\xrightarrow{d}N(\mathbf0,\H^{-1}\J\H^{-1}),
\end{align*}
as $K\to\infty$.
Furthermore, the sandwich variance estimator satisfies
\begin{align*}
K\widehat\V_{\mathrm{sand}}&\xrightarrow{p}\H^{-1}\J\H^{-1} \mathrm{~as~} K\to\infty.
\end{align*}
\end{theorem}

Theorem~\ref{thm:asymptotic-normality} establishes consistency and asymptotic normality for the fixed-basis pseudo-true value and consistency of its variance estimator, supporting target-time interval estimation. Conditions and proof are given in Appendices B.1 and B.2, respectively. Inference for the true mean function $\mu_g(t)$ must also account for approximation error.

At fixed time $t$, let $\widehat\mu_g(t)=\x_g(t)^\top\widehat\bbe$. Its asymptotic variance, $\x_g(t)^\top\H^{-1}\J\H^{-1}\x_g(t)/K$, is estimated by $\x_g(t)^\top\widehat\V_{\mathrm{sand}}\x_g(t)$, giving the $(1-\alpha)$ Wald confidence interval for $\mu_g^\star(t)$:
\begin{align*}
\widehat\mu_g(t)\ \pm\ z_{1-\alpha/2}
\sqrt{\x_g(t)^\top\widehat\V_{\mathrm{sand}}\x_g(t)},
\end{align*}
where $z_{1-\alpha/2}$ is the $1-\alpha/2$ quantile of the standard normal distribution. For groups $g_1$ and $g_2$, define the estimator of their difference as $\widehat\Delta_\mu(t;g_1,g_2)=\{\x_{g_1}(t)-\x_{g_2}(t)\}^\top\widehat\bbe$ and the target as $\Delta_\mu^\star(t;g_1,g_2)=\mu_{g_1}^\star(t)-\mu_{g_2}^\star(t)$. The asymptotic $(1-\alpha)$ Wald confidence interval for this target is
\begin{align*}
\widehat\Delta_\mu(t;g_1,g_2)\ \pm\ z_{1-\alpha/2}
\sqrt{\x_{g_1-g_2}(t)^\top
\widehat\V_{\mathrm{sand}}
\x_{g_1-g_2}(t)},
\end{align*}
where $\x_{g_1-g_2}(t)=\x_{g_1}(t)-\x_{g_2}(t)$. For few studies, small-sample corrections with studies as clusters can be used \citep{pustejovsky2018small}, including CR2 correction and $t$ intervals based on Satterthwaite degrees of freedom. Details of the small-sample correction are given in Appendix A.

For prediction intervals for the latent function of a new study, we use the approach of \citet{higgins2009re}. For the new study's latent function excluding the offset, $f_{\mathrm{new},g}(t)=\x_g(t)^\top\bbe+\z_g(t)^\top\u_{\mathrm{new}}$, a pointwise $(1-\alpha)$ normal-approximation prediction interval is

\begin{align*}
\widehat\mu_g(t)\ \pm\ z_{1-\alpha/2}
\sqrt{\x_g(t)^\top\widehat\V_{\mathrm{sand}}\x_g(t)
+\z_g(t)^\top\widehat\D\z_g(t)}.
\end{align*}

For the between-group difference in the new study, we replace the interval center by $\widehat\Delta_\mu(t;g_1,g_2)$ and replace $\x_g(t)$ and $\z_g(t)$ by $\x_{g_1-g_2}(t)$ and $\z_{g_1-g_2}(t)=\z_{g_1}(t)-\z_{g_2}(t)$, respectively.

\section{Simulation study}
\label{sec:simulation}

To investigate the estimation properties of the proposed method, we compared it with existing methods using data generated under PH and NPH.

\subsection{Primary research questions and design}
\label{sec:simulation-methods}

\paragraph{Data-generating mechanisms}
We evaluated the gain in estimation precision from curve information in studies that did not directly report the target time, and the finite-sample performance and limitations of time-specific inference under PH and NPH. We generated data for $K$ independent studies, with the Negative group denoted by $g=0$ and the Positive group by $g=1$, and 150 participants per group. We divided time into intervals $I_j=(12(j-1),12j]$ months ($j=1,2,3$), and defined the hazard for group $g$ in study $k$ and the survival function for a typical study as

\begin{align*}
 h_{kg}(t)&=\lambda_0 m_{gj}\exp(b_k+g u_k),\qquad t\in I_j,
 \\
 S_g^\star(t)&=\exp\!\left\{-\lambda_0\sum_{j=1}^{3}m_{gj}\ell_j(t)\right\},\qquad
 \ell_j(t)=\max\{0,\min(t,12j)-12(j-1)\}.
\end{align*}

\begin{figure}[htbp]
    \centering
    \includegraphics[width=1.0\linewidth]{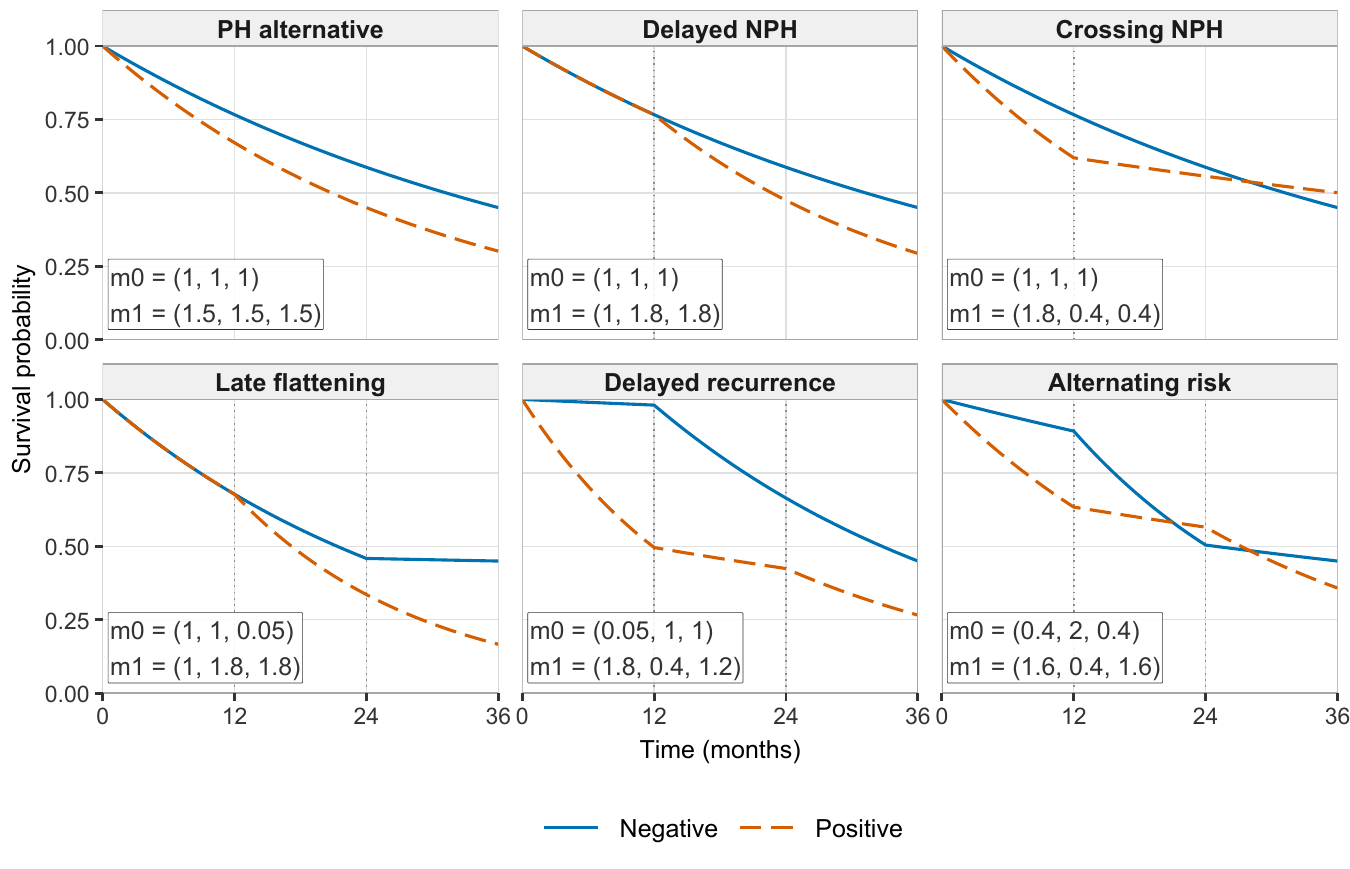}
    \caption{Data-generating mechanisms in the simulation. Each panel shows the group-specific survival curves for a typical study; $m_0$ and $m_1$ indicate the interval-specific hazard multipliers for the Negative and Positive groups. Vertical dotted lines mark the times at which the hazards change.}
    \label{fig:simulation-design}
\end{figure}

The independent variables $b_k\sim N(0,0.30^2)$ and $u_k\sim N(0,0.20^2)$ represented, respectively, between-study heterogeneity in baseline risk common to both groups and in the group effect. We set $\lambda_0=-\log(0.45)/(12\sum_{j=1}^{3}m_{0j})$ so that $S_0^\star(36)=0.45$. Figure~\ref{fig:simulation-design} shows the survival curves and interval-specific hazard multipliers $m_{gj}$ for the six data-generating mechanisms; the curve shapes described below refer to a typical study. The PH alternative has a constant hazard ratio of 1.5. In Delayed NPH, there is no early between-group difference, which emerges after 12 months. In Crossing NPH, the hazard ordering reverses at 12 months, and the survival curves cross subsequently. In Late flattening, the hazard in the Negative group decreases after 24 months; in Delayed recurrence, its early hazard is lowered and then increases after 12 months. In Alternating risk, the hazard ordering between the groups switches at 12 and 24 months.

The primary analysis used $K=10$ studies and the additional analysis used $K=20$. Each of the six data-generating mechanisms was combined with the following three information settings, with 10,000 replicates attempted per combination. Independent exponentially distributed censoring times were calibrated so that, averaged across both groups in a typical study, the probability of censoring before 36 months was 20\%. Administrative censoring was applied at 36 months.

\paragraph{Information settings}
In C1+D1, a randomly selected half of the studies provided C1 data from 0--36 months, and the remainder provided D1 data at 36 months. C1-short+D1 likewise assigned a randomly selected half of the studies to provide curves and the remainder to provide D1 at 36 months, but the curves were observed only through 24 months; this setting evaluated information use from studies that did not directly report the target time. In C2+D1, half the studies provided C2 and the remainder provided D1 at 36 months, evaluating the combination of curves with sparse risk-set information. Risk sets for C2 were reported at 12-month intervals, and the same information format was assigned to both groups within a study.

\paragraph{Comparator methods}
The proposed method used 12 basis functions per group and random intercepts with group-specific variances for between-study random effects. The offset parameters were $t_0=1$ month and $\rho=1$. We also evaluated Proposed-L0, which fixes the smoothing parameter at $\lambda=0$. Interval estimation used the CR2--Satterthwaite correction.

As existing methods, we applied the two-stage Cox and Arends methods to every data-generating mechanism and information setting at both study counts. For the two-stage Cox method, study-specific log hazard ratios estimated from the original individual-level data were pooled using a random-effects model with between-study variance estimated by REML. We constructed 95\% confidence intervals using the Hartung--Knapp--Sidik--Jonkman (HKSJ) method \citep{knapp2003improved,sidik2002simple}.

As a comparator for curve meta-analysis, we used equation (7) of \citet{arends2008}. Group, a quadratic polynomial in log time, and interactions between group and each time term were fixed effects; a study-specific random intercept shared by both groups was the random effect. C1/C2 survival probabilities at monthly time points within the observed range and D1 survival probabilities at reported time points were entered into the REML estimation. Although \citet{arends2008} used model-based Wald intervals, we additionally evaluated Arends-SW, which uses CR2--Satterthwaite intervals to align its inference with the HKSJ method for the two-stage Cox analysis and the sandwich variance of the proposed method. The primary performance measures were root mean squared error (RMSE) and coverage probability of 95\% confidence interval, evaluated for all methods with respect to the true 36-month log cumulative hazard ratio. We calculated them from successful estimation replicates and tabulated estimation failures separately. Numerical performance summaries and additional results for $K=20$ are given in Appendix D.

\subsection{Primary results}

\begin{figure}[htbp]
    \centering
    \includegraphics[width=1.0\linewidth]{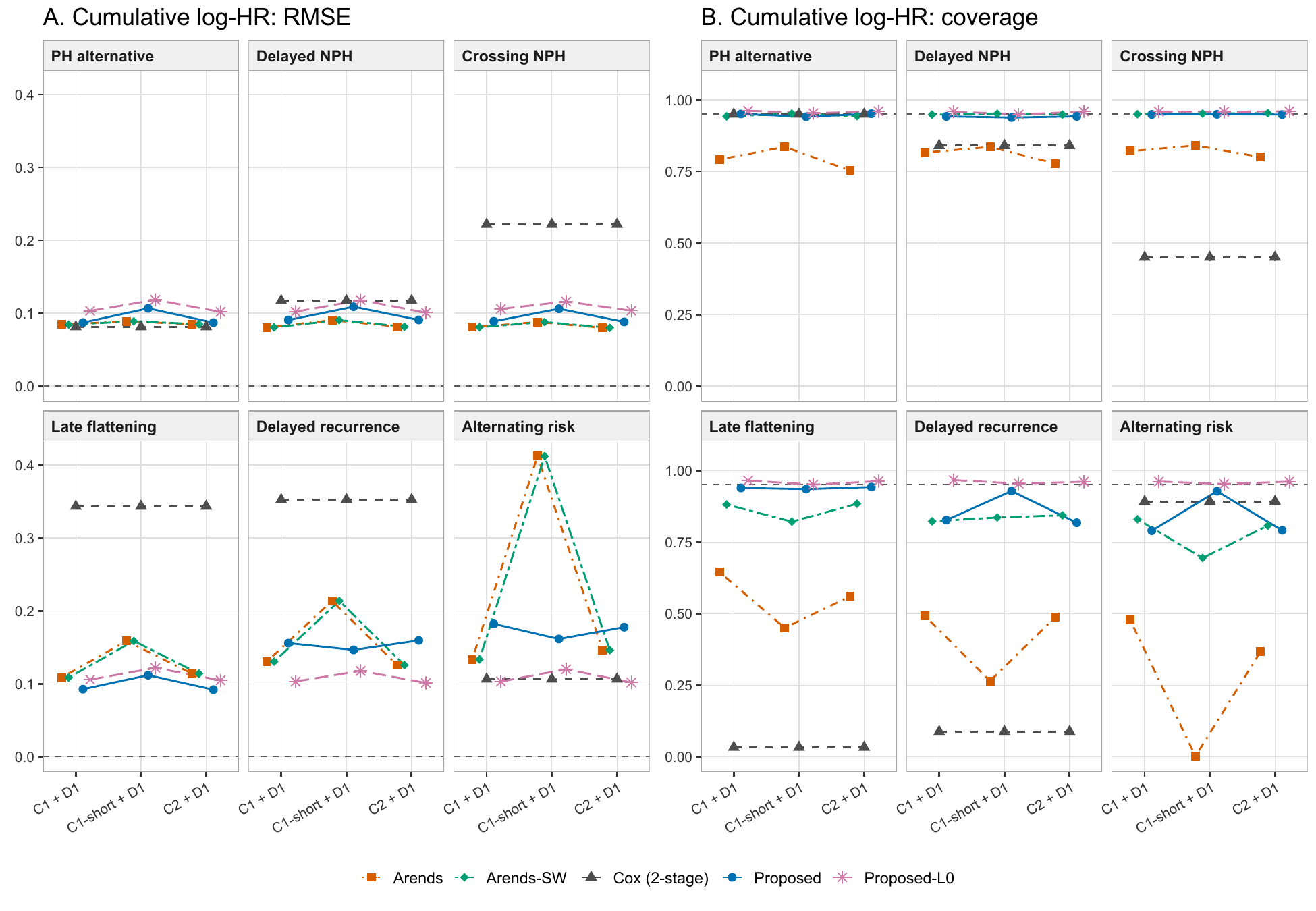}
    \caption{Simulation results for the 36-month log cumulative hazard ratio with $K=10$. Panel A shows RMSE, and Panel B shows coverage probability of 95\% confidence intervals. The horizontal dashed line in Panel B indicates nominal coverage, and vertical lines indicate Monte Carlo 95\% intervals for the coverage probabilities.}
    \label{fig:simulation-primary}
\end{figure}

Figure~\ref{fig:simulation-primary} shows the RMSE and coverage probability of the 36-month log cumulative hazard ratio for the six data-generating mechanisms and three information settings with $K=10$. The Cox method performed differently under PH and NPH. In Delayed NPH and Crossing NPH, the proposed method had a lower RMSE and coverage close to the nominal level. Because the Cox coefficient differs from the log cumulative hazard ratio being evaluated, bias under NPH also affected interval estimation for the Cox method. Under the PH alternative, the Cox method had a lower RMSE, and both methods had coverage close to the nominal level. In these three mechanisms, the Arends method had a lower RMSE than the proposed method, but its model-based Wald intervals showed undercoverage. Arends-SW applied the CR2 correction to the same point estimates, producing wider intervals and coverage closer to the nominal level.

In Late flattening, Delayed recurrence, and Alternating risk, performance varied by curve shape and information setting. For Delayed recurrence and Alternating risk under C1-short+D1, the proposed method had a lower RMSE than Arends-SW and coverage closer to the nominal level. Under C1+D1 and C2+D1 for those same mechanisms, its RMSE was higher than that of the Arends method and it showed clear undercoverage. The proposed method also showed somewhat low coverage in Late flattening. Arends-SW had higher coverage than the Arends method in all three mechanisms, but remained below the nominal level.

The unpenalized Proposed-L0 also had coverage close to the nominal level in Late flattening, Delayed recurrence, and Alternating risk, but had the most numerical optimization failures associated with the absence of smoothing. Its RMSE was lower than that of the proposed method in Delayed recurrence and Alternating risk, but higher in Late flattening. Additional results for a larger number of studies are presented in Appendix D.

\section{Real-data analysis}
\label{sec:rda}

\paragraph{Data and analysis methods}
We analyzed DFS data by baseline ctDNA status from Section~\ref{sec:motivating-example} \citep{chen2023prognostic}, treating 10 of the 12 studies as C1 and the remaining 2 as C2. The analysis covered 0--48 months, with 36 months as the primary time point and 12, 24, and 48 months as secondary time points. We estimated group-specific meta-analytic survival curves and cumulative hazard ratios (CHRs). For the proposed method, the Arends method, Arends-SW, and Proposed-L0, we used the common specifications in Section~\ref{sec:simulation-methods}, with an analysis period of 0--48 months. We also report KM curves obtained by simply pooling patients and published Cox hazard ratios pooled with HKSJ intervals \citep{viechtbauer2010metafor,knapp2003improved,sidik2002simple}. Sources of the data and published estimates, and details of comparator methods and settings, are given in Appendix C.

\paragraph{Primary results}

\begin{figure}[htbp]
    \centering
    \includegraphics[width=1.0\linewidth]{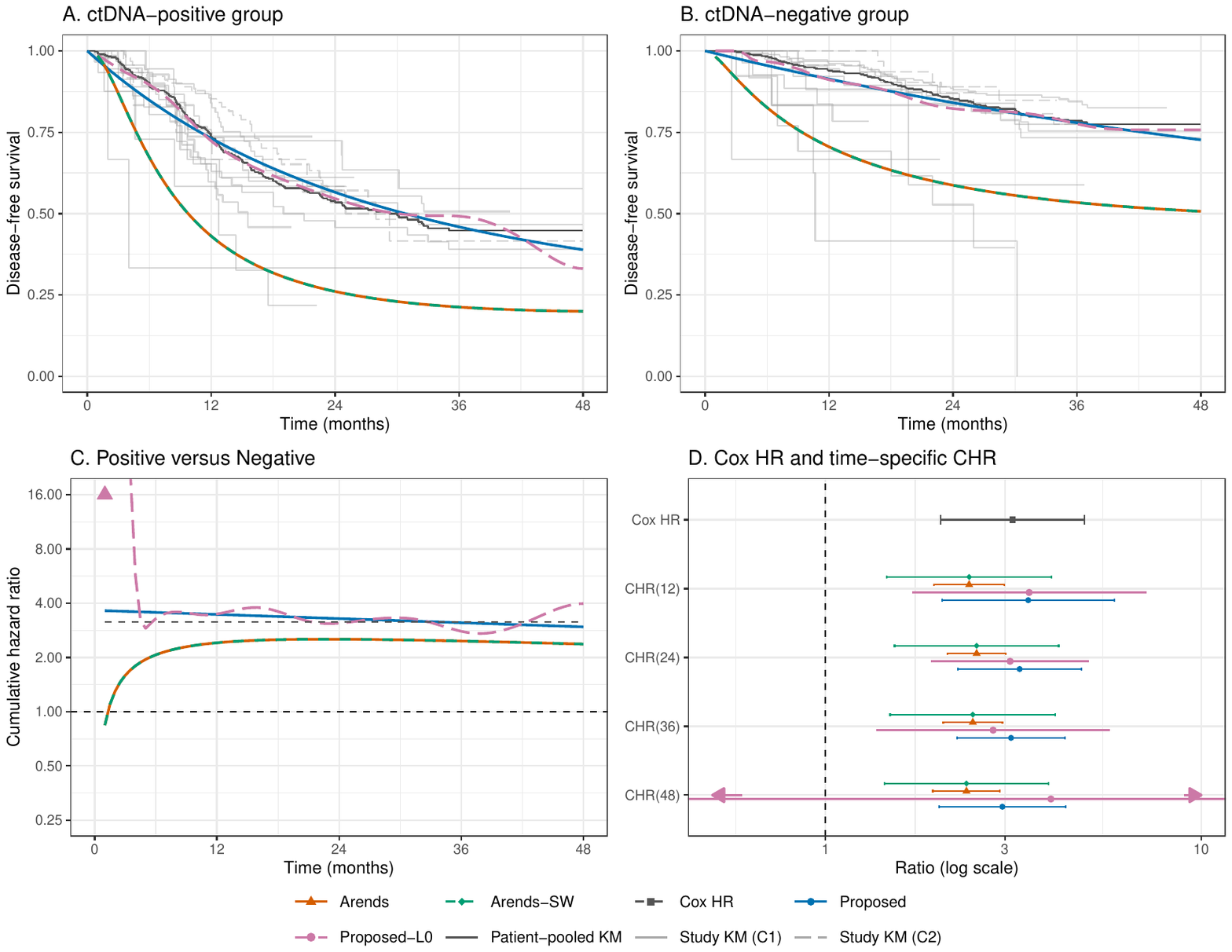}
    \caption{Comparison of DFS by ctDNA status and between-group effects. Panels A--B show the survival curves for the Positive and Negative groups, respectively; Panel C shows the trajectory of the Positive-versus-Negative CHR and the published Cox HR. Panel D shows CHRs at the target times and the published Cox HR, with 95\% confidence intervals. Arrows indicate intervals extending beyond the displayed range.}
    \label{fig:realdata-primary}
\end{figure}

The proposed method estimated lower survival probabilities and higher cumulative hazards in the Positive group than in the Negative group (Figure~\ref{fig:realdata-primary}). The Arends method estimated lower survival probabilities in both groups than the proposed method.

Interval widths varied with the variance estimation method and smoothing choice. The Arends method and Arends-SW gave identical CHR point estimates at each target time, while the intervals from Arends-SW, which uses study-level CR2 correction, were wider (Figure~\ref{fig:realdata-primary}D). The 36-month CHR from Proposed-L0 showed the same direction of the between-group difference as the proposed method but had a wider interval; at 48 months, its interval expanded markedly. Thus, interval estimation near the end of follow-up was sensitive to the smoothing choice.

\section{Discussion}
\label{sec:discussion}

The idea of jointly synthesizing survival probabilities at multiple times is shared with the random-effects model of \citet{arends2008}, but their method was not optimized for combining studies with dense and sparse time points. By treating C1, C2, and D1 as common functional data, the proposed method allows inference without constructing or updating a sampling correlation matrix across times.

Evaluating the proposed method requires consideration of both the estimand and curve shape. Under PH, the Cox and proposed methods target a common between-group effect; under NPH, their different estimands---the Cox coefficient and the time-specific log cumulative hazard ratio---affect the comparison. Smoothing supports estimation stability but can introduce bias for some curve shapes and affect coverage. The improved coverage of unpenalized estimation in the simulations and the wider intervals near the end of follow-up in the application highlight the importance of assessing the degree of smoothing in terms of both curve fit and estimation stability.

The theory developed here concerns the pseudo-true value of a working model with a fixed basis. Inference for the true mean function requires joint evaluation of basis approximation error and stochastic coefficient-estimation error. Extending the theory to allow the basis dimension to increase with the number of studies, while accounting for smoothing and working-covariance estimation, remains future work within the framework of sieve $M$-estimation \citep{chen2007large}. We used CR2--Satterthwaite inference for few studies \citep{pustejovsky2018small}, but this correction does not eliminate bias from mean-function approximation or smoothing. Our intervals are pointwise and hold the selected smoothing parameter, variance components, and working covariance fixed. If working-precision estimation has a first-order effect under mean-structure misspecification, variance evaluation requires estimating equations for variance components and related quantities. Despite these remaining issues, the proposed method provides a useful way to link survival information from studies with different reporting formats and observation periods to comparisons at a prespecified target time.

\section*{Funding}
This work was supported by JSPS KAKENHI Grant Number JP26K21184.

\section*{Supplementary Material}
The appendices, supplementary tables and figures, descriptions of the data sources referenced, and analysis code in Sections 2--5 are included in Supplementary Material. 

\bibliographystyle{biom}
\bibliography{references}
\clearpage
\phantomsection
\addcontentsline{toc}{section}{Supplementary Material}
\begin{center}
\LARGE Supplementary Material for\\[0.5em]
Functional meta-analysis of survival curves across reporting formats and time points
\end{center}
\begin{center}
\large Keisuke Hanada
\end{center}

\appendix
\setcounter{figure}{0}
\setcounter{table}{0}
\setcounter{equation}{0}
\setcounter{theorem}{0}
\renewcommand{\thefigure}{S\arabic{figure}}
\renewcommand{\thetable}{S\arabic{table}}
\renewcommand{\theequation}{S\arabic{equation}}
\renewcommand{\thetheorem}{S\arabic{theorem}}
\renewcommand{\theHfigure}{supp.\arabic{figure}}
\renewcommand{\theHtable}{supp.\arabic{table}}
\renewcommand{\theHequation}{supp.\arabic{equation}}
\renewcommand{\theHtheorem}{supp.\arabic{theorem}}

These supplementary materials provide details supporting the methods and analyses in the main text. Appendix A describes interval estimation using CR2, Appendix B gives the conditions and proof of Theorem 1, Appendix C describes the data sources and analysis settings for the application, and Appendix D presents numerical tables and additional simulation results.

\section{Interval estimation using CR2}
\label{app:estimation}

\label{app:cr2}
We apply the CR2 correction and contrast-specific Satterthwaite degrees of freedom to the study-level sandwich variance estimator in equation~\eqref{eq:cluster-sandwich} of the main text~\citep{pustejovsky2018small}. Using $\widehat\A$ from that equation, let $\widehat\Q_k=\V_k(\widehat\D)^{-1}$. Let $\X$ denote the design matrix obtained by stacking all studies, and define
\[
\widehat\Q=\operatorname{blockdiag}_k(\widehat\Q_k),\qquad
\widehat\V=\operatorname{blockdiag}_k\{\V_k(\widehat\D)\}.
\]
Let $\R=\I-\X\widehat\A^{-1}\X^\top\widehat\Q$. Let $\C_k$ be the diagonal block for study $k$ of $\R\widehat\V\R^\top$, and use an inverse square-root adjustment satisfying $\G_k\C_k\G_k^\top=\V_k(\widehat\D)$ on the estimable subspace. With residuals $\widehat\r_k=\y_k^{(\rho)}-\X_k\widehat\bbe$, the CR2-adjusted coefficient variance estimator is
\begin{align*}
\widehat\V_{\mathrm{CR2}}
&=\widehat\A^{-1}
\left\{\sum_{k=1}^K
\X_k^\top\widehat\Q_k\G_k\widehat\r_k
\widehat\r_k^\top\G_k^\top\widehat\Q_k\X_k
\right\}
\widehat\A^{-1}.
\end{align*}
The term in braces is the sum, across studies, of the outer products of the CR2-adjusted study-level scores $\X_k^\top\widehat\Q_k\G_k\widehat\r_k$.

The degrees of freedom are calculated separately for each contrast, holding the selected smoothing parameter and working covariance fixed. Let $\C_{k\ell}$ denote the block of $\R\widehat\V\R^\top$ corresponding to studies $k$ and $\ell$ ($\C_{kk}=\C_k$). For a contrast vector $\a$, define
\begin{align*}
\boldsymbol h_k(\a)&=\G_k^\top\widehat\Q_k\X_k\widehat\A^{-1}\a,\\
\boldsymbol M(\a)&=\left[\boldsymbol h_k(\a)^\top\C_{k\ell}\boldsymbol h_\ell(\a)\right]_{k,\ell=1}^K.
\end{align*}
Approximating the variance estimator by a quadratic form in a centered normal vector gives mean $\operatorname{tr}\{\boldsymbol M(\a)\}$ and variance $2\operatorname{tr}\{\boldsymbol M(\a)^2\}$. Matching these to the moments of a scaled chi-squared distribution gives the Satterthwaite degrees of freedom \citep{pustejovsky2018small}
\begin{align*}
\nu(\a)
&=\frac{\left[\operatorname{tr}\{\boldsymbol M(\a)\}\right]^2}
{\operatorname{tr}\{\boldsymbol M(\a)^2\}}
=\frac{\left\{\sum_{k=1}^K M_{kk}(\a)\right\}^2}
{\sum_{k=1}^K\sum_{\ell=1}^K M_{k\ell}(\a)^2}.
\end{align*}
Here, $M_{k\ell}(\a)$ is the $(k,\ell)$ element of $\boldsymbol M(\a)$.

A pointwise $(1-\alpha)$ confidence interval for the group-specific mean function $\mu_g^\star(t)$ at a fixed time $t$ is
\begin{align*}
\widehat\mu_g(t)\ \pm\ t_{\nu_g(t),\,1-\alpha/2}
\sqrt{\x_g(t)^\top\widehat\V_{\mathrm{CR2}}\x_g(t)}.
\end{align*}
Here, $\nu_g(t)=\nu\{\x_g(t)\}$, and $t_{\nu,1-\alpha/2}$ is the $1-\alpha/2$ quantile of the $t$ distribution with $\nu$ degrees of freedom.

A pointwise $(1-\alpha)$ confidence interval for the between-group difference $\Delta_\mu^\star(t;g_1,g_2)$ is
\begin{align*}
\widehat\Delta_\mu(t;g_1,g_2)\ \pm\ t_{\nu_{g_1-g_2}(t),\,1-\alpha/2}
\sqrt{\x_{g_1-g_2}(t)^\top
\widehat\V_{\mathrm{CR2}}
\x_{g_1-g_2}(t)}.
\end{align*}
Here, $\x_{g_1-g_2}(t)=\x_{g_1}(t)-\x_{g_2}(t)$ and $\nu_{g_1-g_2}(t)=\nu\{\x_{g_1-g_2}(t)\}$. Exponentiating both endpoints of this interval gives a confidence interval for the cumulative hazard ratio.

\section{Conditions and proof for the asymptotic theory}
\label{app:theory}

\subsection{Regularity conditions and scope}
\label{sup2:sec:regularity}
We use the pseudo-true value $\bbe^\star$, scores $\boldsymbol\psi_k^\star$, $\H$, and $\J$ defined in Section 3.4 of the main text. The basis, number of groups $G$, analysis domain $[0,\tau]$, known offset, and positive semidefinite penalty $\P$ are fixed. The asymptotic variable is the number of independent studies $K$. In the proof, we abbreviate the population quadratic criterion minimized in equation~\eqref{eq:pseudo-true-target} of the main text as $M(\bbe)$.

The study-level data $\mathcal Z_k$ include the information format, reporting times, time-width weights, design matrix, marginal variances, and $\y_k^{(\rho)}$. Let $\V_k^\star$ denote the working matrix corresponding to a fixed population value of the nuisance parameters, and define
\[
\Q_k^\star=(\V_k^\star)^{-1},\qquad
\r_k(\bbe)=\y_k^{(\rho)}-\X_k\bbe,\qquad
\boldsymbol B_k^\star=\X_k^\top\Q_k^\star\X_k.
\]
The working covariance matrix $\V_k^\star$ depends on the study-level data. The marginal variances estimated within each study and their estimation errors are included in the distribution of $\mathcal Z_k$.

We use the Euclidean norm for vectors and the Frobenius norm for matrices below.
\begin{enumerate}
\item[(i)] The complete study-level data $\mathcal Z_1,\ldots,\mathcal Z_K$ are independent and identically distributed.
The matrix $\Q_k^\star$ is symmetric positive definite almost surely, and
$M$ is finite in an open neighborhood of $\bbe^\star$.
\item[(ii)] The matrix $\H$ is positive definite, and
\[
E\|\boldsymbol B_k^\star\|^2<\infty,
\qquad E\|\boldsymbol\psi_k^\star\|^2<\infty.
\]
The matrix $\J$ is positive semidefinite. The asymptotic variance of each contrast for which a Wald interval is constructed is positive.
\item[(iii)] On an event whose probability tends to one, all $\widehat\Q_k$ are symmetric positive definite. Define
\[
\widehat{\boldsymbol B}_k=\X_k^\top\widehat\Q_k\X_k,
\qquad
\boldsymbol d_k=\X_k^\top(\widehat\Q_k-\Q_k^\star)\r_k(\bbe^\star).
\]
The following four conditions hold under the joint distribution:
\begin{align}
K^{-1/2}\sum_{k=1}^K\boldsymbol d_k&=o_p(1),
\label{sup2:eq:precision-first-order}\\
K^{-1}\sum_{k=1}^K
\|\widehat{\boldsymbol B}_k-\boldsymbol B_k^\star\|&=o_p(1),
\label{sup2:eq:hessian-perturbation}\\
K^{-1}\sum_{k=1}^K\|\boldsymbol d_k\|^2&=o_p(1),
\label{sup2:eq:score-L2}\\
K^{-1}\sum_{k=1}^K\|\widehat{\boldsymbol B}_k\|^2&=O_p(1).
\label{sup2:eq:hessian-L2}
\end{align}
These conditions are imposed on the joint distribution, including the dependence between the estimated working precision and the response.
\item[(iv)] We have $\widehat\lambda\geq0$ and
$\widehat\lambda/\sqrt K\xrightarrow{p}0$.
If shape constraints are imposed, minimization is over a closed convex set
$\mathcal C$ defined by a fixed finite collection of linear inequalities, and $\bbe^\star$ is an interior point of $\mathcal C$.
That is, for some $\epsilon>0$,
$\{\bbe:\|\bbe-\bbe^\star\|<\epsilon\}\subset\mathcal C$.
For unconstrained estimation, $\mathcal C=\mathbb R^{p_1}$.
\end{enumerate}
Condition~\eqref{sup2:eq:precision-first-order} requires the first-order contribution of the estimated working precision to the coefficient score to be $o_p(1)$ and is imposed on the joint distribution of the response and the estimated working precision.
When monotonicity constraints are included, the constraint set in condition (iv) is defined by nonnegativity constraints on a fixed finite grid $\mathcal G\cap(0,\tau]$. Each constraint expression is strictly positive at $\bbe^\star$.
The assumption of independent and identically distributed observations applies to the study-level data, including the reporting process. The pseudo-true value $\bbe^\star$ is determined by the working covariance $\V_k^\star$ and depends on its choice under misspecification of the mean structure. In a correctly specified model with mean coefficient $\bbe_0$, the condition $E\{\X_k^\top\Q_k^\star\r_k(\bbe_0)\}=\mathbf0$ implies $\bbe^\star=\bbe_0$. When the precision matrix is constructed from the same data as the response, this condition is likewise imposed on their joint distribution.

\subsection{Proof of Theorem 1 in the main text}
\label{sup2:sec:main-theorem-proof}

\begin{proof}
All vector and matrix dimensions are fixed. Under conditions (i) and (ii), for any fixed vector $\boldsymbol h$ and real number $s$,
\begin{align*}
M(\bbe^\star+s\boldsymbol h)-M(\bbe^\star)
&=-2s\boldsymbol h^\top E(\boldsymbol\psi_k^\star)
  +s^2\boldsymbol h^\top\H\boldsymbol h
\end{align*}
Since $\bbe^\star$ minimizes the population criterion, differentiation at $s=0$ gives
\begin{align*}
\boldsymbol h^\top E(\boldsymbol\psi_k^\star)&=0
\quad(\forall\boldsymbol h),
&
E(\boldsymbol\psi_k^\star)&=\mathbf0
\end{align*}
Consequently, by condition (ii),
\begin{align*}
M(\bbe^\star+\boldsymbol h)-M(\bbe^\star)
&=\boldsymbol h^\top\H\boldsymbol h>0
\qquad(\boldsymbol h\ne\mathbf0)
\end{align*}
so $\bbe^\star$ is unique.

Let $\widehat\H_K=\sum_{k=1}^K\widehat{\boldsymbol B}_k$ and
$\widehat\A=\widehat\H_K+\widehat\lambda\P$.
By conditions (i), (ii), (iv), \eqref{sup2:eq:hessian-perturbation}, and the law of large numbers,
\begin{align*}
\left\|K^{-1}\widehat\A-\H\right\|
&\leq
\left\|K^{-1}\sum_{k=1}^K\boldsymbol B_k^\star-\H\right\|
+K^{-1}\sum_{k=1}^K
 \|\widehat{\boldsymbol B}_k-\boldsymbol B_k^\star\|
+\frac{\widehat\lambda}{K}\|\P\|\\
&=o_p(1).
\end{align*}
Positive definiteness of $\H$, together with the symmetry in condition (iii), implies
\begin{align*}
\Pr(\widehat\A\succ0)&\to1,
&
(K^{-1}\widehat\A)^{-1}&\xrightarrow{p}\H^{-1}
\end{align*}

On the event $\widehat\A\succ0$, let $\widetilde\bbe$ denote the unconstrained minimizer.
The normal equation, with the nuisance parameters fixed at their estimated values, gives
\begin{align*}
\widehat\A\widetilde\bbe
&=\sum_{k=1}^K\X_k^\top\widehat\Q_k\y_k^{(\rho)},\\
\widehat\A(\widetilde\bbe-\bbe^\star)
&=\sum_{k=1}^K\X_k^\top\widehat\Q_k\r_k(\bbe^\star)
  -\widehat\lambda\P\bbe^\star\\
&=\sum_{k=1}^K\boldsymbol\psi_k^\star
  +\sum_{k=1}^K\boldsymbol d_k
  -\widehat\lambda\P\bbe^\star
\end{align*}
Conditions \eqref{sup2:eq:precision-first-order} and (iv) imply
\begin{align*}
K^{-1}\sum_{k=1}^K\boldsymbol d_k
&=K^{-1/2}
  \left(K^{-1/2}\sum_{k=1}^K\boldsymbol d_k\right)
  =o_p(K^{-1/2}),\\
\frac{\widehat\lambda}{K}\P\bbe^\star
&=K^{-1/2}\frac{\widehat\lambda}{\sqrt K}\P\bbe^\star
  =o_p(K^{-1/2}).
\end{align*}
By conditions (i) and (ii) and the law of large numbers,
\begin{align*}
\widetilde\bbe-\bbe^\star
&=(K^{-1}\widehat\A)^{-1}
  \left\{
  K^{-1}\sum_{k=1}^K\boldsymbol\psi_k^\star
  +K^{-1}\sum_{k=1}^K\boldsymbol d_k
  -\frac{\widehat\lambda}{K}\P\bbe^\star
  \right\}
  \xrightarrow{p}\mathbf0.
\end{align*}

Also, by conditions (i) and (ii) and the multivariate central limit theorem,
\begin{align*}
K^{-1/2}\sum_{k=1}^K\boldsymbol\psi_k^\star
&\xrightarrow{d}N(\mathbf0,\J).
\end{align*}
Using \eqref{sup2:eq:precision-first-order}, condition (iv), and Slutsky's theorem,
\begin{align}
\sqrt K(\widetilde\bbe-\bbe^\star)
&=(K^{-1}\widehat\A)^{-1}
  \left\{
  K^{-1/2}\sum_{k=1}^K\boldsymbol\psi_k^\star
  +K^{-1/2}\sum_{k=1}^K\boldsymbol d_k
  -\frac{\widehat\lambda}{\sqrt K}\P\bbe^\star
  \right\}
  \label{sup2:eq:normal-equation-expansion}\\
&=(K^{-1}\widehat\A)^{-1}
  \left\{
  K^{-1/2}\sum_{k=1}^K\boldsymbol\psi_k^\star+o_p(1)
  \right\}
  \notag\\
&\xrightarrow{d}
  N(\mathbf0,\H^{-1}\J\H^{-1}).
  \notag
\end{align}

Under condition (iv), choose $\epsilon>0$ such that
$\{\bbe:\|\bbe-\bbe^\star\|<\epsilon\}\subset\mathcal C$.
Consistency and strict convexity of the quadratic criterion imply
\begin{align*}
\Pr(\widehat\bbe=\widetilde\bbe)
&\geq
\Pr\!\left(
\|\widetilde\bbe-\bbe^\star\|<\epsilon,\,
\widehat\A\succ0
\right)
\to1.
\end{align*}
Hence,
\begin{align*}
\widehat\bbe&\xrightarrow{p}\bbe^\star,
&
\sqrt K(\widehat\bbe-\bbe^\star)
&\xrightarrow{d}N(\mathbf0,\H^{-1}\J\H^{-1}).
\end{align*}

For the variance estimator, let
$\widehat{\boldsymbol\psi}_k=\X_k^\top\widehat\Q_k\r_k(\widehat\bbe)$ and
$\boldsymbol e_k=\widehat{\boldsymbol\psi}_k-\boldsymbol\psi_k^\star$.
Then
\begin{align*}
\boldsymbol e_k
&=\X_k^\top\widehat\Q_k\r_k(\widehat\bbe)
  -\X_k^\top\Q_k^\star\r_k(\bbe^\star)\\
&=\boldsymbol d_k
  -\widehat{\boldsymbol B}_k(\widehat\bbe-\bbe^\star).
\end{align*}
By consistency and conditions \eqref{sup2:eq:score-L2} and \eqref{sup2:eq:hessian-L2},
\begin{align*}
E_K
&:=K^{-1}\sum_{k=1}^K\|\boldsymbol e_k\|^2\\
&\leq
2K^{-1}\sum_{k=1}^K\|\boldsymbol d_k\|^2
+2\|\widehat\bbe-\bbe^\star\|^2
 K^{-1}\sum_{k=1}^K\|\widehat{\boldsymbol B}_k\|^2\\
&=o_p(1).
\end{align*}
Furthermore, by conditions (i) and (ii) and the law of large numbers,
\begin{align*}
T_K
&:=K^{-1}\sum_{k=1}^K\|\boldsymbol\psi_k^\star\|^2
  \xrightarrow{p}E\|\boldsymbol\psi_k^\star\|^2<\infty,\\
K^{-1}\sum_{k=1}^K
\boldsymbol\psi_k^\star\boldsymbol\psi_k^{\star\top}
&\xrightarrow{p}\J.
\end{align*}
The triangle and Cauchy--Schwarz inequalities give
\begin{align*}
&\left\|
K^{-1}\sum_{k=1}^K
\left(
\widehat{\boldsymbol\psi}_k\widehat{\boldsymbol\psi}_k^\top
-\boldsymbol\psi_k^\star\boldsymbol\psi_k^{\star\top}
\right)
\right\|\\
&\quad=
\left\|
K^{-1}\sum_{k=1}^K
\left(
\boldsymbol e_k\boldsymbol\psi_k^{\star\top}
+\boldsymbol\psi_k^\star\boldsymbol e_k^\top
+\boldsymbol e_k\boldsymbol e_k^\top
\right)
\right\|\\
&\quad\leq
2K^{-1}\sum_{k=1}^K
\|\boldsymbol e_k\|\|\boldsymbol\psi_k^\star\|
+E_K\\
&\quad\leq 2\sqrt{E_KT_K}+E_K
=o_p(1).
\end{align*}
It follows that
\begin{align*}
K^{-1}\sum_{k=1}^K
\widehat{\boldsymbol\psi}_k\widehat{\boldsymbol\psi}_k^\top
&\xrightarrow{p}\J,\\
K\widehat\V_{\mathrm{sand}}
&=(K^{-1}\widehat\A)^{-1}
  \left(
  K^{-1}\sum_{k=1}^K
  \widehat{\boldsymbol\psi}_k\widehat{\boldsymbol\psi}_k^\top
  \right)
  (K^{-1}\widehat\A)^{-1}\\
&\xrightarrow{p}\H^{-1}\J\H^{-1}.
\end{align*}
\end{proof}

\section{Supplementary details for the application}
\label{app:realdata}

\subsection{Data used in the application}

Data for 13 studies were obtained from the publicly available supplementary materials of \citet{chen2023prognostic}. One study had no DFS events among patients with observed baseline ctDNA status and was excluded, leaving 12 studies for analysis. For 10 studies, Kaplan--Meier curves and Greenwood variances were constructed from patient-level supplementary data and treated as C1. For the remaining two studies, reconstructed survival data were used, but risk-set information was restricted to that reported in the original publications to construct C2 inputs. Risk sets were available at month 0 for the study by Zhang et al. and at months 0, 10, 20, 30, 40, and 50 for the study by Provencio et al. Curve inputs were restricted to the intersection of the 0--48-month analysis domain and the observed range of each study and group. For C2 curve drops after the final reported risk-set time, the reference variance was also computed by carrying forward the most recently reported risk-set count.

Published hazard ratios in the same materials were matched by study name. When multiple endpoints were listed, the first listed DFS, PFS, or recurrence endpoint was used. From the published estimate $\widehat{\mathrm{HR}}_k$ and its 95\% confidence interval $[L_k,U_k]$, the log-scale estimate and standard error were calculated as $y_k=\log\widehat{\mathrm{HR}}_k$ and $s_k=\{\log U_k-\log L_k\}/(2z_{0.975})$. A REML random-effects model was fitted using $s_k^2$ as the within-study variance, and an HKSJ interval was constructed~\citep{viechtbauer2010metafor,knapp2003improved,sidik2002simple}. Table~\ref{s2:tab:reported-hr} presents the extracted values.

\begin{table}[htbp]
\centering
\caption{Published hazard ratios for the 12 studies included in the analysis, obtained from the supplementary materials of \citet{chen2023prognostic}, and log-scale standard errors calculated from the reported 95\% confidence intervals.}
\label{s2:tab:reported-hr}
\resizebox{\textwidth}{!}{\input{tables/table_reported_hr_estimates}}
\par\smallskip\footnotesize
REML pooled HR: 3.141 (95\% confidence interval 2.022--4.880), $\widehat\tau_{\mathrm{HR}}^2=0.298$.
\end{table}

\subsection{Details of the Arends method}

To implement the Arends comparison described in Section 2.3 of the main text, we used equation (7) of the original article by Arends et al.~\citep{arends2008}. For the group indicator $g\in\{0,1\}$, we specified
\begin{align*}
 \log\{-\log\widehat S_{kg}(t)\}
 ={}&\alpha_0+\alpha_1g+(\alpha_2+\alpha_3g)\log t
 +(\alpha_4+\alpha_5g)(\log t)^2+u_k+\varepsilon_{kg}(t),\\
 &u_k\sim N(0,\sigma_u^2).
\end{align*}
A study-specific random intercept shared by both groups was estimated by REML. The C1/C2 curves were evaluated right-continuously at monthly times within their observed ranges, retaining observations satisfying $0<\widehat S<1$. Reference Greenwood variances were used for C2, and the transformed marginal standard errors were held fixed. Within the same study and group, for $t_a<t_b$, the correlation was specified as
\begin{align*}
 R_{ab}=\left[\frac{S_{kg}(t_b)\{1-S_{kg}(t_a)\}}
 {\{1-S_{kg}(t_b)\}S_{kg}(t_a)}\right]^{1/2},
\end{align*}
and updated iteratively using survival probabilities predicted from the fixed effects. The sampling-error covariance between different groups was set to zero.

The Arends method used model-based normal Wald intervals for the fixed effects. Arends-SW used the same point estimates and iterative updates of the within-study correlations as the Arends method, but applied study-cluster CR2 correction and $t$ intervals with target-specific Satterthwaite degrees of freedom. Intervals for survival probabilities and CHRs were obtained by back-transformation from the transformed scale.
For both methods, fixed-effect standard errors were calculated using the iteratively updated within-study correlations and the REML estimate of the between-study variance.

\section{Supplementary simulation results}
\label{app:simulation}

\subsection{Main results for the six data-generating mechanisms}

Table~\ref{s2:tab:simulation-main} presents the performance measures and numbers of successful fits for $K=10$, corresponding to Figure 3 in the main text.

\begingroup
\footnotesize
\input{tables/table_simulation_main_supplement}
\endgroup

\subsection{Additional comparisons with more studies}

\begin{figure}[H]
    \centering
    \includegraphics[width=1.0\linewidth]{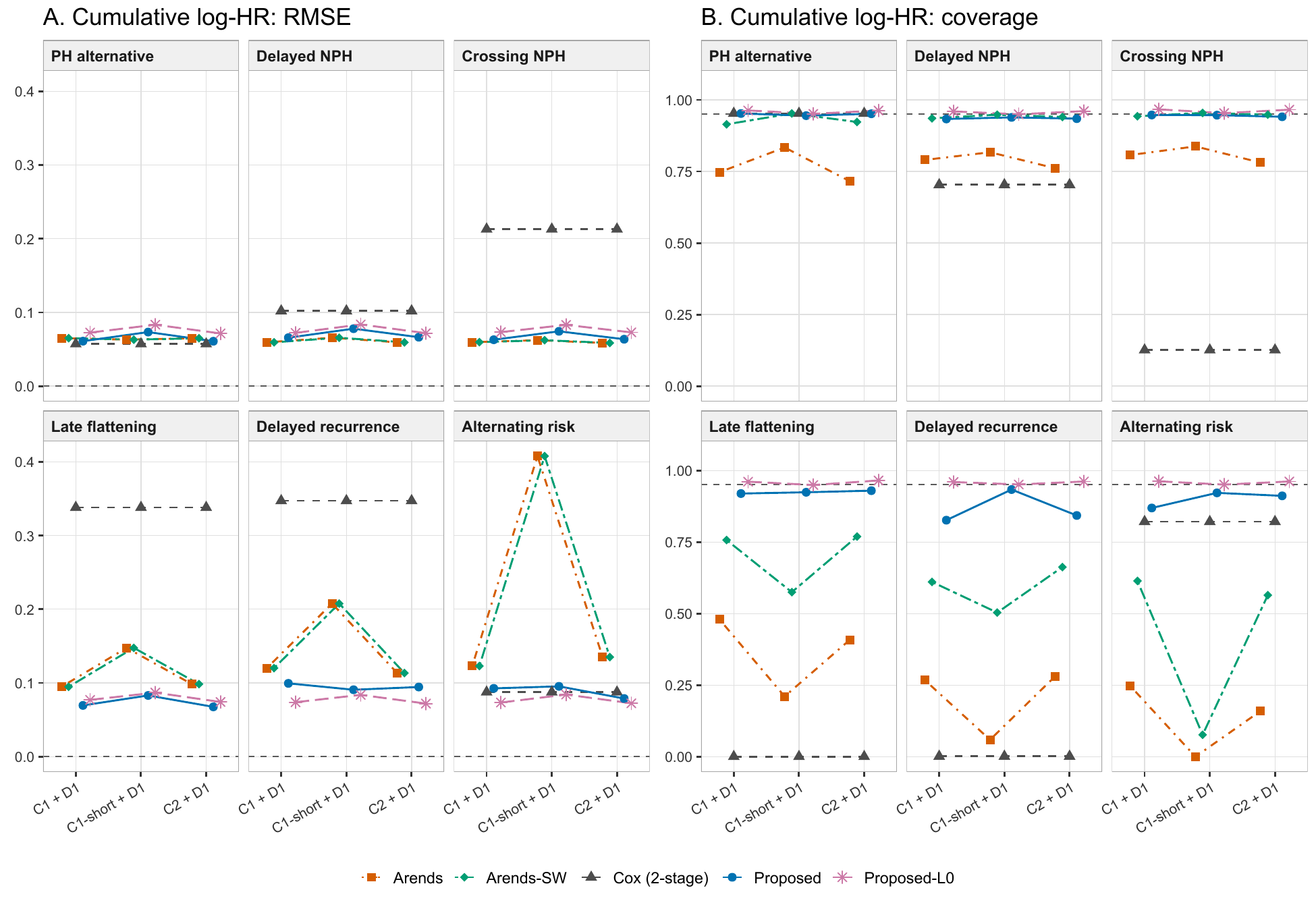}
    \caption{Simulation results for the log cumulative hazard ratio at 36 months with $K=20$. Panel A shows RMSE, and Panel B shows coverage probability of 95\% confidence intervals. In Panel B, horizontal dashed lines indicate nominal coverage, and vertical lines indicate Monte Carlo 95\% intervals for the coverage probabilities.}
    \label{s2:fig:simulation-k20}
\end{figure}

Results for $K=20$ are presented in Figure~\ref{s2:fig:simulation-k20} and Table~\ref{s2:tab:simulation-k20}. Compared with $K=10$, RMSE and mean interval width decreased for every method and setting. Undercoverage of the Arends method and Arends-SW became more pronounced under Late flattening, Delayed recurrence, and Alternating risk. The proposed method's RMSE also decreased under Delayed recurrence and Alternating risk, but undercoverage remained under C1 + D1 and C2 + D1 for both mechanisms. Proposed-L0 had coverage probabilities close to the nominal level across all mechanisms and information settings. As for $K=10$, optimization nonconvergence was concentrated under C1-short + D1. Estimation failures of the Arends method and Arends-SW occurred in the same replicates.

\begingroup
\footnotesize
\input{tables/table_simulation_k20_supplement}
\endgroup
\end{document}

%% file: def.tex
\def\bbe{{\text{\boldmath $\beta$}}}

\def\a{{\text{\boldmath $a$}}}

\def\r{{\text{\boldmath $r$}}}

\def\u{{\text{\boldmath $u$}}}

\def\x{{\text{\boldmath $x$}}}
\def\y{{\text{\boldmath $y$}}}
\def\z{{\text{\boldmath $z$}}}
\def\A{{\text{\boldmath $A$}}}

\def\C{{\text{\boldmath $C$}}}
\def\D{{\text{\boldmath $D$}}}

\def\G{{\text{\boldmath $G$}}}
\def\H{{\text{\boldmath $H$}}}
\def\I{{\text{\boldmath $I$}}}
\def\J{{\text{\boldmath $J$}}}

\def\P{{\text{\boldmath $P$}}}
\def\Q{{\text{\boldmath $Q$}}}
\def\R{{\text{\boldmath $R$}}}

\def\V{{\text{\boldmath $V$}}}
\def\W{{\text{\boldmath $W$}}}
\def\X{{\text{\boldmath $X$}}}

\def\Z{{\text{\boldmath $Z$}}}

%% file: tables/table_reported_hr_estimates.tex
\begin{tabular}{lllrr}
\toprule
Study & Information class & Reported HR [95\% CI] & log HR & Reconstructed SE \\
\midrule
Zhang JT et al. (2022) & C2 & 15.25 [7.95, 29.27] & 2.7246 & 0.3325 \\
Yue D et al. (2022) & C1 & 1.03 [0.13, 8.42] & 0.0296 & 1.0640 \\
Xia L et al. (2021) & C1 & 2.60 [1.30, 5.10] & 0.9555 & 0.3487 \\
Waldeck S et al. (2021) & C1 & 2.01 [0.41, 9.89] & 0.6981 & 0.8120 \\
Qiu B et al. (2021) & C1 & 2.64 [1.02, 6.83] & 0.9708 & 0.4851 \\
Provencio M et al. (2022) & C2 & 3.75 [1.31, 10.72] & 1.3218 & 0.5363 \\
Peng M et al. (2020) & C1 & 3.65 [1.84, 7.22] & 1.2945 & 0.3482 \\
Ohara S et al. (2020) & C1 & 2.76 [0.46, 16.55] & 1.0152 & 0.9140 \\
Li N et al. (2022) & C1 & 2.42 [1.11, 5.27] & 0.8838 & 0.3974 \\
Kuang PP et al. (2020) & C1 & 1.92 [0.45, 8.21] & 0.6523 & 0.7408 \\
Gale D et al. (2022) & C1 & 3.14 [1.49, 6.60] & 1.1442 & 0.3797 \\
Abbosh C et al. (2017) & C1 & 1.27 [0.35, 4.65] & 0.2390 & 0.6599 \\
\bottomrule
\end{tabular}

%% file: tables/table_simulation_main_supplement.tex
\footnotesize
\setlength{\tabcolsep}{3pt}
\begin{longtable}{@{}lllrrrrr@{}}
\caption{Simulation results for the log cumulative hazard ratio at 36 months. Each setting used 10,000 replicates, and performance measures are conditional on successful fits. Interval width is the mean width of the 95\% confidence intervals.}\label{s2:tab:simulation-main}\\
\toprule
DGP & Format & Method & Success/total & Bias & RMSE & Coverage & Mean width\\
\midrule
\endfirsthead
\toprule
DGP & Format & Method & Success/total & Bias & RMSE & Coverage & Mean width\\
\midrule
\endhead
\bottomrule
\endfoot
PH alternative & C1 + D1 & Arends & 10000/10000 & -0.037 & 0.084 & 0.792 & 0.214 \\*
 &  & Arends-SW & 10000/10000 & -0.037 & 0.084 & 0.942 & 0.352 \\*
 &  & Cox (2-stage) & 10000/10000 & 0.004 & 0.081 & 0.950 & 0.357 \\*
 &  & Proposed & 10000/10000 & 0.002 & 0.087 & 0.951 & 0.383 \\*
 &  & Proposed-L0 & 10000/10000 & 0.004 & 0.103 & 0.962 & 0.559 \\
\addlinespace[1pt]
 & C1-short + D1 & Arends & 10000/10000 & -0.011 & 0.089 & 0.836 & 0.247 \\*
 &  & Arends-SW & 10000/10000 & -0.011 & 0.089 & 0.952 & 0.401 \\*
 &  & Cox (2-stage) & 10000/10000 & 0.004 & 0.081 & 0.950 & 0.357 \\*
 &  & Proposed & 10000/10000 & -0.002 & 0.107 & 0.941 & 0.483 \\*
 &  & Proposed-L0 & 9041/10000 & 0.001 & 0.118 & 0.953 & 0.622 \\
\addlinespace[1pt]
 & C2 + D1 & Arends & 9999/10000 & -0.035 & 0.085 & 0.753 & 0.199 \\*
 &  & Arends-SW & 9999/10000 & -0.035 & 0.085 & 0.943 & 0.360 \\*
 &  & Cox (2-stage) & 10000/10000 & 0.004 & 0.081 & 0.950 & 0.357 \\*
 &  & Proposed & 10000/10000 & 0.003 & 0.087 & 0.952 & 0.383 \\*
 &  & Proposed-L0 & 10000/10000 & 0.004 & 0.102 & 0.960 & 0.548 \\
\midrule
Delayed NPH & C1 + D1 & Arends & 10000/10000 & -0.026 & 0.081 & 0.816 & 0.213 \\*
 &  & Arends-SW & 10000/10000 & -0.026 & 0.081 & 0.948 & 0.352 \\*
 &  & Cox (2-stage) & 10000/10000 & -0.086 & 0.117 & 0.840 & 0.351 \\*
 &  & Proposed & 10000/10000 & 0.024 & 0.091 & 0.942 & 0.386 \\*
 &  & Proposed-L0 & 10000/10000 & 0.001 & 0.102 & 0.959 & 0.556 \\
\addlinespace[1pt]
 & C1-short + D1 & Arends & 10000/10000 & 0.018 & 0.091 & 0.836 & 0.249 \\*
 &  & Arends-SW & 10000/10000 & 0.018 & 0.091 & 0.952 & 0.407 \\*
 &  & Cox (2-stage) & 10000/10000 & -0.086 & 0.117 & 0.840 & 0.351 \\*
 &  & Proposed & 10000/10000 & 0.021 & 0.109 & 0.938 & 0.494 \\*
 &  & Proposed-L0 & 8997/10000 & -0.000 & 0.118 & 0.949 & 0.620 \\
\addlinespace[1pt]
 & C2 + D1 & Arends & 10000/10000 & -0.023 & 0.082 & 0.778 & 0.198 \\*
 &  & Arends-SW & 10000/10000 & -0.023 & 0.082 & 0.949 & 0.364 \\*
 &  & Cox (2-stage) & 10000/10000 & -0.086 & 0.117 & 0.840 & 0.351 \\*
 &  & Proposed & 10000/10000 & 0.027 & 0.091 & 0.942 & 0.386 \\*
 &  & Proposed-L0 & 10000/10000 & 0.001 & 0.101 & 0.959 & 0.541 \\
\midrule
Crossing NPH & C1 + D1 & Arends & 9997/10000 & 0.025 & 0.081 & 0.821 & 0.220 \\*
 &  & Arends-SW & 9997/10000 & 0.025 & 0.081 & 0.950 & 0.353 \\*
 &  & Cox (2-stage) & 10000/10000 & 0.204 & 0.222 & 0.450 & 0.383 \\*
 &  & Proposed & 10000/10000 & -0.006 & 0.089 & 0.949 & 0.389 \\*
 &  & Proposed-L0 & 10000/10000 & 0.014 & 0.106 & 0.959 & 0.582 \\
\addlinespace[1pt]
 & C1-short + D1 & Arends & 9998/10000 & -0.006 & 0.088 & 0.841 & 0.247 \\*
 &  & Arends-SW & 9998/10000 & -0.006 & 0.088 & 0.952 & 0.399 \\*
 &  & Cox (2-stage) & 10000/10000 & 0.204 & 0.222 & 0.450 & 0.383 \\*
 &  & Proposed & 10000/10000 & 0.013 & 0.106 & 0.950 & 0.495 \\*
 &  & Proposed-L0 & 9016/10000 & 0.012 & 0.116 & 0.958 & 0.634 \\
\addlinespace[1pt]
 & C2 + D1 & Arends & 9996/10000 & 0.021 & 0.080 & 0.800 & 0.206 \\*
 &  & Arends-SW & 9996/10000 & 0.021 & 0.080 & 0.953 & 0.356 \\*
 &  & Cox (2-stage) & 10000/10000 & 0.204 & 0.222 & 0.450 & 0.383 \\*
 &  & Proposed & 10000/10000 & -0.010 & 0.088 & 0.949 & 0.386 \\*
 &  & Proposed-L0 & 10000/10000 & 0.012 & 0.103 & 0.960 & 0.575 \\
\midrule
Late flattening & C1 + D1 & Arends & 9997/10000 & -0.078 & 0.108 & 0.645 & 0.214 \\*
 &  & Arends-SW & 9997/10000 & -0.078 & 0.108 & 0.881 & 0.356 \\*
 &  & Cox (2-stage) & 10000/10000 & -0.335 & 0.343 & 0.032 & 0.340 \\*
 &  & Proposed & 10000/10000 & -0.031 & 0.093 & 0.939 & 0.390 \\*
 &  & Proposed-L0 & 10000/10000 & -0.024 & 0.106 & 0.965 & 0.594 \\
\addlinespace[1pt]
 & C1-short + D1 & Arends & 9999/10000 & -0.134 & 0.159 & 0.450 & 0.246 \\*
 &  & Arends-SW & 9999/10000 & -0.134 & 0.159 & 0.821 & 0.453 \\*
 &  & Cox (2-stage) & 10000/10000 & -0.335 & 0.343 & 0.032 & 0.340 \\*
 &  & Proposed & 10000/10000 & -0.035 & 0.112 & 0.935 & 0.495 \\*
 &  & Proposed-L0 & 9180/10000 & -0.023 & 0.122 & 0.950 & 0.639 \\
\addlinespace[1pt]
 & C2 + D1 & Arends & 9993/10000 & -0.080 & 0.114 & 0.560 & 0.192 \\*
 &  & Arends-SW & 9993/10000 & -0.080 & 0.114 & 0.884 & 0.381 \\*
 &  & Cox (2-stage) & 10000/10000 & -0.335 & 0.343 & 0.032 & 0.340 \\*
 &  & Proposed & 10000/10000 & -0.029 & 0.092 & 0.942 & 0.390 \\*
 &  & Proposed-L0 & 9997/10000 & -0.023 & 0.105 & 0.962 & 0.577 \\
\midrule
Delayed recurrence & C1 + D1 & Arends & 9997/10000 & -0.107 & 0.130 & 0.492 & 0.213 \\*
 &  & Arends-SW & 9997/10000 & -0.107 & 0.130 & 0.822 & 0.376 \\*
 &  & Cox (2-stage) & 10000/10000 & 0.341 & 0.353 & 0.087 & 0.405 \\*
 &  & Proposed & 10000/10000 & -0.123 & 0.156 & 0.827 & 0.456 \\*
 &  & Proposed-L0 & 9984/10000 & 0.004 & 0.103 & 0.966 & 0.758 \\
\addlinespace[1pt]
 & C1-short + D1 & Arends & 9987/10000 & -0.191 & 0.214 & 0.265 & 0.263 \\*
 &  & Arends-SW & 9987/10000 & -0.191 & 0.214 & 0.836 & 0.612 \\*
 &  & Cox (2-stage) & 10000/10000 & 0.341 & 0.353 & 0.087 & 0.405 \\*
 &  & Proposed & 10000/10000 & -0.078 & 0.146 & 0.928 & 0.599 \\*
 &  & Proposed-L0 & 8896/10000 & 0.002 & 0.118 & 0.954 & 0.625 \\
\addlinespace[1pt]
 & C2 + D1 & Arends & 9996/10000 & -0.099 & 0.125 & 0.487 & 0.194 \\*
 &  & Arends-SW & 9996/10000 & -0.099 & 0.125 & 0.843 & 0.374 \\*
 &  & Cox (2-stage) & 10000/10000 & 0.341 & 0.353 & 0.087 & 0.405 \\*
 &  & Proposed & 10000/10000 & -0.128 & 0.159 & 0.817 & 0.449 \\*
 &  & Proposed-L0 & 9985/10000 & 0.002 & 0.101 & 0.961 & 0.788 \\
\midrule
Alternating risk & C1 + D1 & Arends & 10000/10000 & -0.110 & 0.133 & 0.477 & 0.212 \\*
 &  & Arends-SW & 10000/10000 & -0.110 & 0.133 & 0.830 & 0.389 \\*
 &  & Cox (2-stage) & 10000/10000 & 0.064 & 0.106 & 0.892 & 0.371 \\*
 &  & Proposed & 10000/10000 & -0.145 & 0.182 & 0.789 & 0.480 \\*
 &  & Proposed-L0 & 10000/10000 & -0.008 & 0.103 & 0.961 & 0.563 \\
\addlinespace[1pt]
 & C1-short + D1 & Arends & 10000/10000 & -0.401 & 0.412 & 0.001 & 0.250 \\*
 &  & Arends-SW & 10000/10000 & -0.401 & 0.412 & 0.694 & 0.927 \\*
 &  & Cox (2-stage) & 10000/10000 & 0.064 & 0.106 & 0.892 & 0.371 \\*
 &  & Proposed & 10000/10000 & -0.087 & 0.161 & 0.927 & 0.632 \\*
 &  & Proposed-L0 & 9044/10000 & -0.007 & 0.120 & 0.953 & 0.626 \\
\addlinespace[1pt]
 & C2 + D1 & Arends & 10000/10000 & -0.123 & 0.146 & 0.367 & 0.193 \\*
 &  & Arends-SW & 10000/10000 & -0.123 & 0.146 & 0.807 & 0.403 \\*
 &  & Cox (2-stage) & 10000/10000 & 0.064 & 0.106 & 0.892 & 0.371 \\*
 &  & Proposed & 10000/10000 & -0.134 & 0.178 & 0.791 & 0.464 \\*
 &  & Proposed-L0 & 10000/10000 & -0.007 & 0.102 & 0.961 & 0.548 \\
\end{longtable}
\normalsize
\setlength{\tabcolsep}{6pt}

%% file: tables/table_simulation_k20_supplement.tex
\footnotesize
\setlength{\tabcolsep}{3pt}
\begin{longtable}{@{}lllrrrrr@{}}
\caption{Simulation results for the log cumulative hazard ratio at 36 months with $K=20$. Each setting used 10,000 replicates, and performance measures are conditional on successful fits.}\label{s2:tab:simulation-k20}\\
\toprule
DGP & Format & Method & Success/total & Bias & RMSE & Coverage & Mean width\\
\midrule
\endfirsthead
\toprule
DGP & Format & Method & Success/total & Bias & RMSE & Coverage & Mean width\\
\midrule
\endhead
\bottomrule
\endfoot
PH alternative & C1 + D1 & Arends & 10000/10000 & -0.037 & 0.065 & 0.747 & 0.151 \\*
 &  & Arends-SW & 10000/10000 & -0.037 & 0.065 & 0.915 & 0.233 \\*
 &  & Cox (2-stage) & 10000/10000 & 0.004 & 0.057 & 0.953 & 0.238 \\*
 &  & Proposed & 10000/10000 & -0.001 & 0.061 & 0.953 & 0.254 \\*
 &  & Proposed-L0 & 10000/10000 & 0.002 & 0.073 & 0.963 & 0.344 \\
\addlinespace[1pt]
 & C1-short + D1 & Arends & 10000/10000 & -0.012 & 0.063 & 0.834 & 0.174 \\*
 &  & Arends-SW & 10000/10000 & -0.012 & 0.063 & 0.953 & 0.265 \\*
 &  & Cox (2-stage) & 10000/10000 & 0.004 & 0.057 & 0.953 & 0.238 \\*
 &  & Proposed & 10000/10000 & -0.003 & 0.073 & 0.945 & 0.310 \\*
 &  & Proposed-L0 & 9609/10000 & -0.001 & 0.083 & 0.951 & 0.371 \\
\addlinespace[1pt]
 & C2 + D1 & Arends & 10000/10000 & -0.035 & 0.065 & 0.716 & 0.141 \\*
 &  & Arends-SW & 10000/10000 & -0.035 & 0.065 & 0.923 & 0.239 \\*
 &  & Cox (2-stage) & 10000/10000 & 0.004 & 0.057 & 0.953 & 0.238 \\*
 &  & Proposed & 10000/10000 & -0.000 & 0.061 & 0.952 & 0.253 \\*
 &  & Proposed-L0 & 10000/10000 & 0.001 & 0.071 & 0.963 & 0.336 \\
\midrule
Delayed NPH & C1 + D1 & Arends & 10000/10000 & -0.025 & 0.059 & 0.790 & 0.151 \\*
 &  & Arends-SW & 10000/10000 & -0.025 & 0.059 & 0.936 & 0.233 \\*
 &  & Cox (2-stage) & 10000/10000 & -0.085 & 0.102 & 0.704 & 0.233 \\*
 &  & Proposed & 10000/10000 & 0.024 & 0.066 & 0.933 & 0.254 \\*
 &  & Proposed-L0 & 9998/10000 & 0.000 & 0.072 & 0.960 & 0.340 \\
\addlinespace[1pt]
 & C1-short + D1 & Arends & 10000/10000 & 0.019 & 0.066 & 0.817 & 0.176 \\*
 &  & Arends-SW & 10000/10000 & 0.019 & 0.066 & 0.948 & 0.268 \\*
 &  & Cox (2-stage) & 10000/10000 & -0.085 & 0.102 & 0.704 & 0.233 \\*
 &  & Proposed & 10000/10000 & 0.022 & 0.078 & 0.939 & 0.316 \\*
 &  & Proposed-L0 & 9498/10000 & -0.001 & 0.084 & 0.950 & 0.369 \\
\addlinespace[1pt]
 & C2 + D1 & Arends & 9999/10000 & -0.021 & 0.059 & 0.761 & 0.140 \\*
 &  & Arends-SW & 9999/10000 & -0.021 & 0.059 & 0.940 & 0.241 \\*
 &  & Cox (2-stage) & 10000/10000 & -0.085 & 0.102 & 0.704 & 0.233 \\*
 &  & Proposed & 10000/10000 & 0.026 & 0.066 & 0.934 & 0.254 \\*
 &  & Proposed-L0 & 10000/10000 & 0.000 & 0.072 & 0.961 & 0.333 \\
\midrule
Crossing NPH & C1 + D1 & Arends & 10000/10000 & 0.025 & 0.060 & 0.807 & 0.155 \\*
 &  & Arends-SW & 10000/10000 & 0.025 & 0.060 & 0.943 & 0.234 \\*
 &  & Cox (2-stage) & 10000/10000 & 0.204 & 0.213 & 0.126 & 0.256 \\*
 &  & Proposed & 10000/10000 & -0.013 & 0.063 & 0.947 & 0.257 \\*
 &  & Proposed-L0 & 10000/10000 & 0.010 & 0.073 & 0.967 & 0.356 \\
\addlinespace[1pt]
 & C1-short + D1 & Arends & 10000/10000 & -0.006 & 0.062 & 0.838 & 0.175 \\*
 &  & Arends-SW & 10000/10000 & -0.006 & 0.062 & 0.955 & 0.264 \\*
 &  & Cox (2-stage) & 10000/10000 & 0.204 & 0.213 & 0.126 & 0.256 \\*
 &  & Proposed & 10000/10000 & 0.005 & 0.074 & 0.947 & 0.314 \\*
 &  & Proposed-L0 & 9560/10000 & 0.005 & 0.083 & 0.954 & 0.375 \\
\addlinespace[1pt]
 & C2 + D1 & Arends & 10000/10000 & 0.021 & 0.059 & 0.781 & 0.145 \\*
 &  & Arends-SW & 10000/10000 & 0.021 & 0.059 & 0.949 & 0.236 \\*
 &  & Cox (2-stage) & 10000/10000 & 0.204 & 0.213 & 0.126 & 0.256 \\*
 &  & Proposed & 10000/10000 & -0.016 & 0.064 & 0.941 & 0.255 \\*
 &  & Proposed-L0 & 10000/10000 & 0.008 & 0.073 & 0.966 & 0.350 \\
\midrule
Late flattening & C1 + D1 & Arends & 10000/10000 & -0.078 & 0.095 & 0.480 & 0.151 \\*
 &  & Arends-SW & 10000/10000 & -0.078 & 0.095 & 0.757 & 0.234 \\*
 &  & Cox (2-stage) & 10000/10000 & -0.334 & 0.338 & 0.000 & 0.224 \\*
 &  & Proposed & 10000/10000 & -0.032 & 0.070 & 0.919 & 0.258 \\*
 &  & Proposed-L0 & 10000/10000 & -0.023 & 0.077 & 0.961 & 0.357 \\
\addlinespace[1pt]
 & C1-short + D1 & Arends & 10000/10000 & -0.135 & 0.148 & 0.210 & 0.174 \\*
 &  & Arends-SW & 10000/10000 & -0.135 & 0.148 & 0.575 & 0.294 \\*
 &  & Cox (2-stage) & 10000/10000 & -0.334 & 0.338 & 0.000 & 0.224 \\*
 &  & Proposed & 10000/10000 & -0.035 & 0.083 & 0.924 & 0.319 \\*
 &  & Proposed-L0 & 9614/10000 & -0.020 & 0.087 & 0.949 & 0.375 \\
\addlinespace[1pt]
 & C2 + D1 & Arends & 10000/10000 & -0.080 & 0.098 & 0.408 & 0.136 \\*
 &  & Arends-SW & 10000/10000 & -0.080 & 0.098 & 0.769 & 0.251 \\*
 &  & Cox (2-stage) & 10000/10000 & -0.334 & 0.338 & 0.000 & 0.224 \\*
 &  & Proposed & 10000/10000 & -0.028 & 0.068 & 0.929 & 0.257 \\*
 &  & Proposed-L0 & 10000/10000 & -0.020 & 0.074 & 0.965 & 0.350 \\
\midrule
Delayed recurrence & C1 + D1 & Arends & 10000/10000 & -0.108 & 0.120 & 0.267 & 0.150 \\*
 &  & Arends-SW & 10000/10000 & -0.108 & 0.120 & 0.611 & 0.248 \\*
 &  & Cox (2-stage) & 10000/10000 & 0.341 & 0.347 & 0.001 & 0.269 \\*
 &  & Proposed & 10000/10000 & -0.057 & 0.099 & 0.826 & 0.280 \\*
 &  & Proposed-L0 & 9998/10000 & 0.001 & 0.074 & 0.959 & 0.356 \\
\addlinespace[1pt]
 & C1-short + D1 & Arends & 9999/10000 & -0.197 & 0.208 & 0.059 & 0.185 \\*
 &  & Arends-SW & 9999/10000 & -0.197 & 0.208 & 0.503 & 0.394 \\*
 &  & Cox (2-stage) & 10000/10000 & 0.341 & 0.347 & 0.001 & 0.269 \\*
 &  & Proposed & 10000/10000 & -0.033 & 0.091 & 0.933 & 0.363 \\*
 &  & Proposed-L0 & 9563/10000 & -0.001 & 0.084 & 0.951 & 0.371 \\
\addlinespace[1pt]
 & C2 + D1 & Arends & 9999/10000 & -0.099 & 0.113 & 0.281 & 0.137 \\*
 &  & Arends-SW & 9999/10000 & -0.099 & 0.113 & 0.663 & 0.248 \\*
 &  & Cox (2-stage) & 10000/10000 & 0.341 & 0.347 & 0.001 & 0.269 \\*
 &  & Proposed & 10000/10000 & -0.047 & 0.094 & 0.843 & 0.273 \\*
 &  & Proposed-L0 & 9998/10000 & 0.001 & 0.072 & 0.961 & 0.362 \\
\midrule
Alternating risk & C1 + D1 & Arends & 9997/10000 & -0.111 & 0.123 & 0.247 & 0.150 \\*
 &  & Arends-SW & 9997/10000 & -0.111 & 0.123 & 0.614 & 0.256 \\*
 &  & Cox (2-stage) & 10000/10000 & 0.064 & 0.088 & 0.821 & 0.246 \\*
 &  & Proposed & 10000/10000 & -0.041 & 0.093 & 0.869 & 0.277 \\*
 &  & Proposed-L0 & 10000/10000 & -0.007 & 0.073 & 0.962 & 0.343 \\
\addlinespace[1pt]
 & C1-short + D1 & Arends & 10000/10000 & -0.402 & 0.408 & 0.000 & 0.176 \\*
 &  & Arends-SW & 10000/10000 & -0.402 & 0.408 & 0.076 & 0.591 \\*
 &  & Cox (2-stage) & 10000/10000 & 0.064 & 0.088 & 0.821 & 0.246 \\*
 &  & Proposed & 10000/10000 & -0.043 & 0.095 & 0.922 & 0.365 \\*
 &  & Proposed-L0 & 9581/10000 & -0.004 & 0.084 & 0.951 & 0.371 \\
\addlinespace[1pt]
 & C2 + D1 & Arends & 10000/10000 & -0.123 & 0.135 & 0.160 & 0.136 \\*
 &  & Arends-SW & 10000/10000 & -0.123 & 0.135 & 0.564 & 0.265 \\*
 &  & Cox (2-stage) & 10000/10000 & 0.064 & 0.088 & 0.821 & 0.246 \\*
 &  & Proposed & 10000/10000 & -0.024 & 0.079 & 0.911 & 0.269 \\*
 &  & Proposed-L0 & 10000/10000 & -0.007 & 0.072 & 0.962 & 0.337 \\
\end{longtable}
\normalsize
\setlength{\tabcolsep}{6pt}